\documentclass[prl,twocolumn,preprintnumbers,superscriptaddress]{revtex4-1}
\usepackage{amssymb}
\usepackage{amsmath}
\usepackage{amsfonts}
\usepackage{wasysym}
\usepackage{graphicx}
\usepackage{dcolumn}
\usepackage{bm}
\usepackage{color}
\usepackage[colorlinks,linkcolor=blue, urlcolor=blue, anchorcolor=blue, citecolor=blue]{hyperref}

\begin{document}

\title{Super-correlated radiance in nonlinear photonic waveguides}

\author{Zhihai Wang}
\affiliation{Center for Quantum Sciences and School of Physics, Northeast Normal University, Changchun 130024, China}
\author{Tuomas Jaako}
\affiliation{Vienna Center for Quantum Science and Technology, Atominstitut, TU Wien, 1040 Vienna, Austria}
\author{Peter Kirton}
\altaffiliation[Present address: ]{Department of Physics and SUPA, University of Strathclyde, Glasgow G4 0NG, UK}
\affiliation{Vienna Center for Quantum Science and Technology, Atominstitut, TU Wien, 1040 Vienna, Austria}
\author{Peter Rabl}
\affiliation{Vienna Center for Quantum Science and Technology, Atominstitut, TU Wien, 1040 Vienna, Austria}

\begin{abstract}
We study the collective decay of two-level emitters coupled to a nonlinear waveguide, for example, a nanophotonic lattice or a superconducting resonator array with strong photon-photon interactions. Under these conditions a new decay channel into bound photon pairs emerges, through which spatial correlations between emitters are established by regular interference as well as interactions between the photons. We derive an effective Markovian theory to model the resulting decay dynamics of an arbitrary distribution of emitters and identify collective effects beyond the usual phenomena of super- and subradiance. Specifically, in the limit of many close-by emitters, we find that the system undergoes a super-correlated decay process where either all the emitters are in the excited state or in the ground state, but not in any of the intermediate states. The predicted effects can be probed in state-of-the-art waveguide QED experiments and provide a striking example of how the dynamics of open quantum systems can be modified by many-body effects in a non-harmonic environment.
\end{abstract}


\maketitle

The radiative decay of an excited atom, induced by its coupling to the continuum of electromagnetic modes, is a prototypical example of irreversible energy loss in quantum systems. Dicke~\cite{Dicke,Gross1982main} showed that this process can be modified significantly in settings with multiple closely spaced emitters, where the decay rate can be collectively enhanced or suppressed due to interference. Recently, such super- and subradiant effects have gained considerable attention in the context of waveguide QED~\cite{Hughes2004,Shen2005,Chang2007,Zhou2008,Longo2010,Zheng2010,Lombardo2014,Shahmoon2016,Roy2017}, where atoms~\cite{ReitzPRL2013,YallaPRL2014,Hood2016,Corzo2019}, quantum dots~\cite{review-lodahl} or superconducting qubits~\cite{Astafiev2010,Hoi2011,Mlynek2014,Mirhosseini2018,Sundaresan2019} are coupled to nanophotonic or microwave waveguides. Along with enhancing the rate of decay, the strong transverse mode confinement in such structures also leads to strongly correlated emission between distant emitters. Under such conditions, collective radiation effects can give rise to self-organization~\cite{ChangPRL2013,GriesserPRL2013}, long-range entanglement~\cite{dzsotjan2010,GonzalesTudela2011,Stannigel2012,Zheng2013,shahmoon2013,Facchi2016} and efficient light-matter interfaces~\cite{AsenjoGarcia2017,Mahmoodian2016}.

Collective radiance is usually modeled under the premise that the environment is represented by independent harmonic oscillators. However, in nanophotonic lattices, plasmonic waveguides and superconducting resonator arrays, intrinsic or engineered nonlinearities~\cite{Hartmann2006,GreentreeNatPhys2006,AngelakisPRA2007,Carusotto2013,Angelakis2017,Houck2012,Degiron2010,Kauranen2012} can become significant at the level of a few photons, breaking the validity of this assumption. Therefore, a natural question arises:  how radiation behaves in a strongly interacting environment. In this Letter, we address this question by analyzing the decay of multiple two-level systems (TLSs) into an array of coupled cavities with strong onsite photon-photon interactions. Specifically, we focus on emitter frequencies below the edge of the propagation band, where single-photon emission is suppressed and an interaction-induced decay channel, {which forces the TLSs to emit photons in bound pairs}, dominates. In this regime the decay dynamics is determined by a new correlation length related to the size of attractively bound photon pairs. These correlations give rise to collectively enhanced and suppressed decay processes beyond the effects of super- and subradiance in linear photonic systems. Most remarkably, for many closely spaced emitters, we find a collective acceleration beyond the $N^2$-scaling of superradiance. This has the intriguing consequence that at any time, almost all TLSs are either found in the excited or the ground state, but not in any of the intermediate mixed configurations. In this limit spontaneous emission becomes perfectly correlated.

\begin{figure}[t]
    \begin{center}
        \includegraphics[width=\columnwidth]{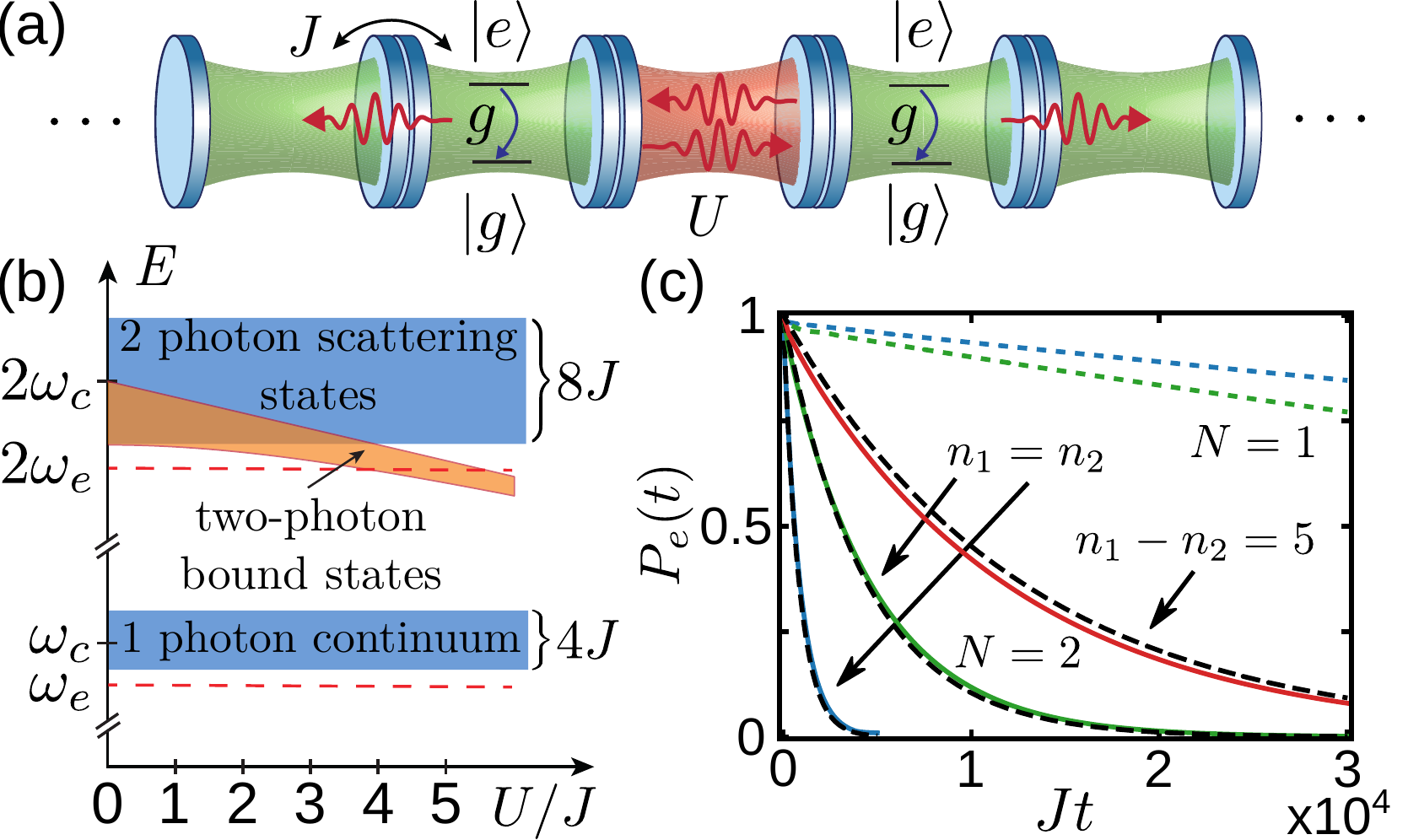}
        \caption{(a) Sketch of the waveguide QED setup, where multiple TLSs are coupled to a photonic lattice with nearest neighbor tunneling $J$ and onsite interaction $U$. (b) The corresponding band structure in the one- and two-photon subspace. (c) Evolution of the excited state population $P_e(t)$ for $N=1$ (dotted) and $N=2$ (solid) emitters with frequency $\omega_e< \omega_c-2J$, as indicated in (b). The parameters for the three solid curves are $U=J$, $K_0\approx0.1\pi$, $g/J=0.02$, $(\omega_e-\omega_c)/J=-2.04$ and $|n_1-n_2|=0(5)$ for the green (red) line and $U=4J$, $K_0\approx0.5\pi$, $g/J=0.1$, $(\omega_e-\omega_c)/J=-2.45$ for the blue line.
        For all plots $\kappa/J=3\times10^{-4}$ and the dashed lines show the approximate analytic result of Eq.~\eqref{approx}.}
        \label{Fig1:Setup}
    \end{center}
\end{figure}

\emph{Model.}---We consider a system  of $N$ TLSs with ground state $|g\rangle$ and excited state  $|e\rangle$, which interact with a one dimensional (1D) array of tunnel-coupled cavities, as schematically shown in Fig.~\ref{Fig1:Setup}(a). The photonic lattice is modeled by a tight-binding Hamiltonian $(\hbar=1)$
\begin{equation}
H_{\rm ph}=\sum_{n}
\omega_c a^\dag_n a_n -  \frac{U}{2} a_n^\dag  a_n^\dag a_n a_
n - J  \left(a_n^\dag a_{n+1}+{\rm H.c.}\right),
\label{Hpn}
\end{equation}
where $a_n$ is the photon annihilation operator on site $n$, and $\omega_c$ and $4J>0$ are the central frequency and  the total width of the propagation band, respectively.
The second term in Eq.~\eqref{Hpn} accounts for a Kerr-like interaction between the photons, which we assume to be attractive, i.e. $U>0$. The Hamiltonian for the whole system is
\begin{equation}\label{eq:Hamiltonian}
H= H_{\rm ph} + \frac{\omega_e}{2} \sum_{i=1}^N \sigma_i^z  + g\sum_{i=1}^N  \left(a_{n_i} \sigma_
{i}^+ + a_{n_i}^\dag \sigma_i^-\right),
\end{equation}
where the $\sigma^{\pm,z}_{i}$ are the usual Pauli operators for the $i$th TLS located at lattice site $n_i$, $\omega_e$ is the TLS transition frequency and $g$ the coupling strength. For small $g$ and $\omega_e \in [\omega_c-2J, \omega_c+2J]$, an excited TLS can decay with a characteristic rate $\Gamma_1\sim  g^2/J$ into a propagating single-photon wavepacket. For multiple TLSs, the emitted photons can interfere, which gives rise to the well-studied effects of super- and subradiance~\cite{Dicke,Gross1982main,SHbook,Black2005,Akkermans2008, Scully2009,Bienaime2012,Ostermann2013,Scully2015,AsenjoGarcia2017,Angerer2018,Zhang2019,Ke2019}.

In the following we are interested in a scenario where  $\omega_e< \omega_c-2J$ lies below the propagation band, such that this regular decay channel is absent. As indicated in Fig.~\ref{Fig1:Setup}(b), under this condition it is still possible for two or more emitters to decay via a resonant excitation of a bound two-photon state. For a lattice of $N_c\gg1$ cavities, {and neglecting the emitters now}, a general two-photon eigenstate can be written as $|\Psi_K\rangle= \frac{1}{\sqrt{2}} \sum_{n,m} \Psi_K(n,m) a^\dag_n a^\dag_m |{\rm vac}\rangle$, where $|{\rm vac}\rangle$ is the vacuum state of the waveguide and the wavefunction, $ \Psi_K(n,m)=e^{iK(n+m)/2}\psi_K(n-m)/\sqrt{N_c}$, is symmetric and can be decomposed into center-of-mass and relative components. For each $K\in (-\pi,\pi]$ there is a band of scattering states, $\psi^q_K(r)\sim \cos(q r-\varphi_K)$, which extend across the whole lattice and have energies $E^q_{K}=2\omega_c-4J_K\cos(q)$, where $J_K=J\cos(K/2)$. In addition, there exists one bound state per $K$ with energy $E_K^b=2\omega_c- \sqrt{U^2+16J_K^2}$~\cite{supp,Piil2007,Valiente2008} and an exponentially localized wavefunction, $\psi^b_K(r)\propto e^{-|r|/\lambda_K}$,
with size $\lambda_K^{-1} =\mathrm{asinh}(\mathcal{U}_K)$, where $\mathcal{U}_K=U/(4J_K)$. As shown in Fig.~\ref{Fig1:Setup}(b), the energies $E_K^b$ largely overlap with the scattering states for small $U$, but for $U\gtrsim J$ a finite band of propagating two-photon states appears below the scattering continuum. The repulsively-bound counterparts of these states have been observed with cold atoms in optical lattices~\cite{Winkler2004} and they also exist in 2D and 3D lattices, although at slightly larger interactions~\cite{supp}.

\begin{figure}[t]
    \begin{center}
        \includegraphics[width=\columnwidth]{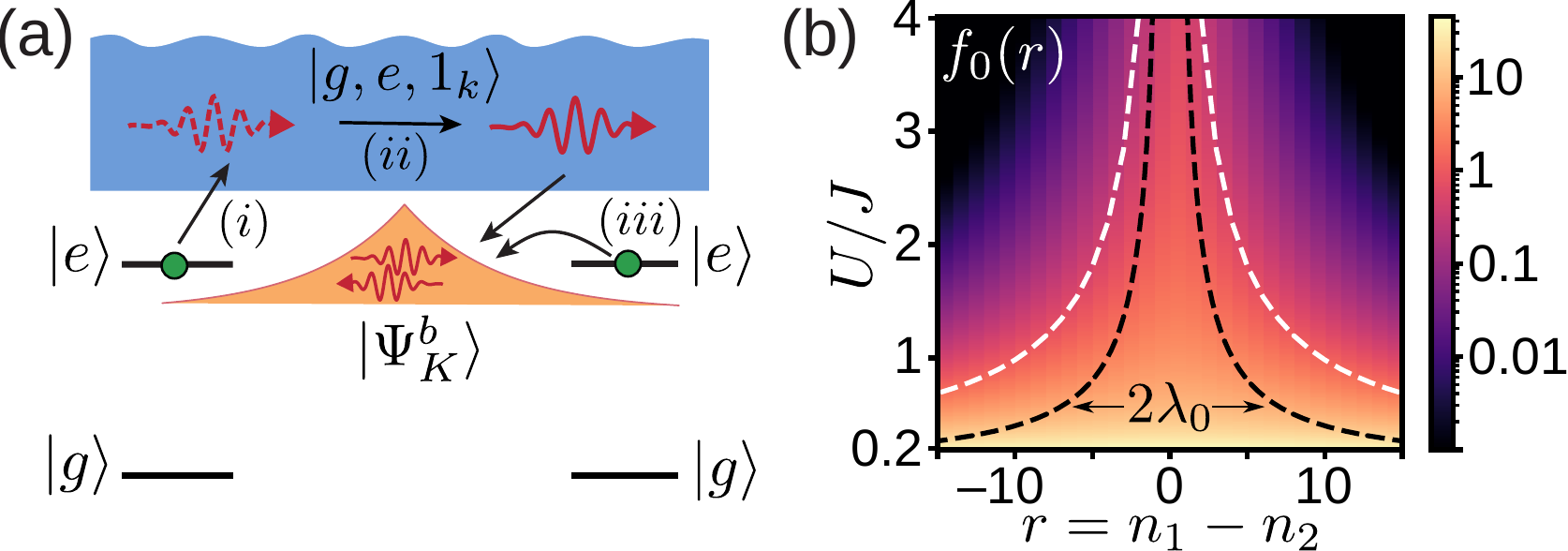}
        \caption{(a) Illustration of the correlated two-photon decay process. The first TLS emits a virtual photon (i), which propagates for a time $\sim 1/\delta_k$ (ii) before it combines with the second emitted photon into a propagating bound state (iii). (b) Contour plot of $f_{K=0}(n_1,n_2)$ for different ratios $U/J$. The black and the white dashed lines represent the bound-state size $\lambda_{K=0}$ and the $1/e$ decay length of $f_0(r)$, respectively.
        }
        \label{Fig2:TwoAtomDecay}
    \end{center}
\end{figure}

\emph{Correlated two-photon decay.}---In Fig.~\ref{Fig1:Setup}(c), we show the evolution of the excited state population, $P_e(t)=\sum_i \langle \sigma_i^+\sigma_i^-\rangle/N$, for both one and two {initially excited} TLSs with a frequency below the band edge and including a small loss rate, $\kappa$, for each cavity. { For a single TLS, we only observe a small residual decay of about $\Gamma_1\approx \kappa g^2/(4\sqrt{J\delta_0^3})$~\cite{supp,Calajo2016}, where $\delta_0=\omega_c-\omega_e-2J$ is the detuning from the band edge}. However, two nearby TLSs decay at a much faster rate, which is approximately independent of $\kappa$. To understand this behavior, we consider the weak-coupling limit $g\ll J,U$ and write the wavefunction of the system as
\begin{equation}
\begin{split}
|\phi_2\rangle(t)= & e^{-2i\omega_{e}t}\Big[c_e(t)\sigma_1^+\sigma_2^+ + \sum_{K}c_{K}(t)B_K^\dag \\
&+ \sum_k ( c_{1k}(t) \sigma_1^+ + c_{2k}(t) \sigma_2^+) a^\dag_k  \Big] |g,g,{\rm vac}\rangle,
\end{split}
\end{equation}
where $a_k^\dag = \sum_n e^{ikn} a^\dag_n/\sqrt{N_c}$ and $B_K^\dag$ is the creation operator for a bound photon pair, $|\Psi^b_K\rangle=B_K^\dag|{\rm vac}\rangle$. This does not include the two-photon scattering states, which play a negligible role in the dynamics~\cite{supp}. Since the one-photon states, $|i,1_k\rangle=\sigma^+_{i}a_k^\dag |g,g,{\rm vac}\rangle$, are separated by an energy gap, { $\delta_k=\omega_c-2J\cos(k)-\omega_e \gg g/\sqrt{N_c}$}, they can be eliminated using perturbation theory. We hence obtain an effective coupling between the TLSs and the continuum of two-photon bound states~\cite{supp},
\begin{eqnarray}
i\dot{c}_e& =& -\frac{g^2}{J\sqrt{N _c}} \sum_{K}   e^{i K(n_1+n_2)/2} f_K(n_1,n_2) c_K,\\
i\dot{c}_K &=& \Delta_K c_{K}-\frac{g^2}{J\sqrt{N _c}}  e^{-i K(n_1+n_2)/2} f_K(n_1,n_2) c_e, \label{eq:dotcK}
\end{eqnarray}
where $\Delta_K= E_{K}^{b}-2\omega_{e}$. {Eq.~\eqref{eq:dotcK} is only valid for bound photon states below the propagation band, $\Delta_K\approx0$, which are, however,  the relevant modes in the following discussion.}  The matrix element $f_K(n_1,n_2)\equiv f_K(n_1-n_2)$ depends only on the relative separation and can be expressed in terms of the two-photon correlation function
\begin{equation}
\begin{split}
f_K(n_1,n_2)&= -i \sqrt{N_c} J \int_0^\infty d\tau \, e^{i K(n_1+n_2)/2}  e^{-i\omega_e \tau}\\
 \times & \langle {\rm vac}| B_K[a^\dag_{n_2} a^\dag_{n_1}(\tau)  +  a^\dag_{n_1}a^\dag_{n_2}(\tau)]|{\rm vac}\rangle.
 \end{split}
\end{equation}
As illustrated in Fig.~\ref{Fig2:TwoAtomDecay}(a), this quantity can be interpreted as follows: The first TLS emits a virtual photon at $n_1$. This photon propagates for a time, $\tau$, before another photon is created by the second emitter at $n_2$. Then $f_K(n_1,n_2)$ is the overlap of this photon pair (and its symmetric counterpart) with the two-photon bound state $|\Psi^b_K\rangle$. Therefore, as shown in Fig.~\ref{Fig2:TwoAtomDecay}(b), the correlations induced can exceed the size of the two-photon bound state, $\lambda_K$, and depend on propagation and interference effects of the intermediate single-photon states.

To proceed we assume that the energy of the emitters lies within the band of bound two-photon states, $2\omega_e\in [E_{0}^b,E_\pi^b]$, and eliminate the dynamics of those states using a Wigner-Weisskopf approximation.  {This is valid when the group velocity of the emitted photons, $v_g(K_0)=\partial E_K^b/\partial K_{K=K_0}$, where $K_0$ is determined by $2\omega_e=E_{K_0}^b$, is large than the effective coupling $\sim g^2/J$}. In this case we obtain an exponential decay of the doubly excited state, $P_e(t)=e^{-\Gamma t}$, with a rate~\cite{supp}
 \begin{equation}
 \Gamma = \frac{2g^4}{J^3} |f_{K_0}(n_{1},n_{2})|^{2} \tilde \rho(K_0),
 \label{approx}
 \end{equation}
where $\tilde \rho(K)=J/v_g(K)$ is the normalized density of bound two-photon states and  $v_g(K)=4J^2\sin(K)/\sqrt{U^2+16J^2\cos^2(K/2)}$.
Fig.~\ref{Fig1:Setup}(c) and additional examples in~\cite{supp} show that this approximate result agrees very well with exact numerical simulations for typical rates in the range of $\Gamma/J\sim 10^{-4}$---$10^{-2}$.

\emph{Collective radiance.}---To analyze the decay of an arbitrary distribution of emitters, we generalize the elimination of the photons from above deriving a master equation (ME)~\cite{BreuerBook} for the reduced density operator, $ \rho $, of the TLSs. In a frame rotating with $\omega_e$, this equation is~\cite{supp}
 \begin{equation}
\dot \rho=	-i\left(H_{\rm eff}\rho-\rho H_{\rm eff}^\dag\right)+\mathcal{J}(\rho),
\label{mastermain}
\end{equation}
where $\mathcal{J}(\rho)=\sum_{i,j,k,l}\Gamma_{ij,kl}\sigma_{i}^{-}\sigma_{j}^{-} \rho\sigma_{k}^{+}\sigma_{l}^{+}$ is the recycling term with $\Gamma_{ij,kl}=\Gamma_0 {\rm Re} \{ A_{ij,kl}\}$ and $\Gamma_0=2g^4\tilde \rho(K_0)/J^3 $ and we have introduced the non-Hermitian Hamiltonian
\begin{equation}
H_{\rm eff}= -i \frac{\Gamma_0}{2}\sum_{i,j,k,l} A_{ij,kl} \sigma_{k}^{+}\sigma_{l}^{+}\sigma_{i}^{-}
 \sigma_{j}^{-}+A_{ij,kl}^*\sigma_{i}^{+}\sigma_{j}^{+}\sigma_{k}^{-}\sigma_{l}^{-},
\end{equation}
which describes collective interactions, $\sim {\rm Im} \{A_{ij,kl}\}$,  and dissipation processes, $\sim {\rm Re} \{ A_{ij,kl}\}$, involving up to four TLSs. The form of the amplitudes
\begin{equation}
A_{ij,kl}= f_{K_{0}}(n_{i},n_{j})f_{K_{0}}(n_{k},n_{l})e^{i K_{0}
 \left|(n_{k}+n_{l})-(n_{i}+n_{j})\right|/2},
\end{equation}
shows that the radiation-induced correlations depend on two processes. First, correlations with a length scale determined by $f_{K_0}$ arise from the nonlinear decay mechanism, as discussed above. Second, photons emitted from different pairs can interfere, which is taken into account by the exponential phase factor. Similar to collective emission in regular waveguides~\cite{Chang2012,GonzalesTudela2013,Calajo2016}, these interference effects are infinite in range, but here they crucially depend on the relative positions of all the TLSs involved.

\begin{figure}[t]
    \begin{center}
        \includegraphics[width=\columnwidth]{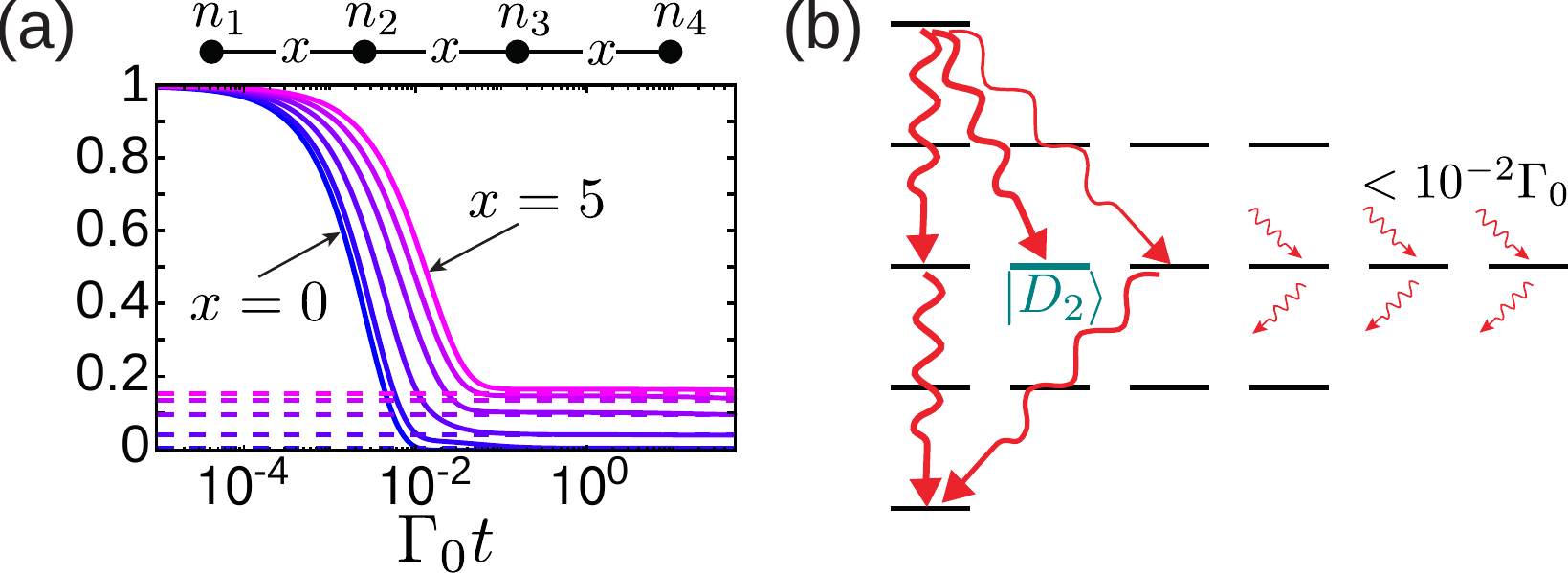}
        \caption{(a) The dynamics of $P_e(t)$, {obtained from Eq.~\eqref{mastermain}}, for $N=4$ TLSs with different spacings, $n_{i+1}-n_i=x$.  The solid lines represent the numerical results and the dashed lines are from a semi-analytic calculation~\cite{supp}. The parameters are  $U=J$, $K_0\approx0.1\pi$, $g/J=0.02$ and $(\omega_e-\omega_c)/J=-2.04$. (b) Sketch of the dominant decay paths into and out of the two-excitation eigenstates of $H_{\rm eff}$. The thickness of the lines indicates the relative strength of the decay rates~\cite{supp}.}
        \label{Fig3:Subradiance}
    \end{center}
\end{figure}

\emph{Subradiance.}---
The coherent and dissipative four-body interactions in $H_{\rm eff}$ make the decay process of a multi-emitter system rather complex and can lead to a speed-up of emission as well as the appearance of subradiance, i.e., weakly or even non-decaying states. All single excitation states, which remain unaffected by two-photon decay, belong to this class of states, but there are additional non-trivial examples. The existence of these states is evident from Fig.~\ref{Fig3:Subradiance}(a), which shows the dynamics of $N=4$ excited TLSs with different spacings $x$ between them. The ME, Eq.~\eqref{mastermain}, ensures that only states with an even number of excitations are populated. We observe a fast initial decay on a timescale $\sim 1/(f^2_{K_0}(x)\Gamma_0)$, after which the system reaches a quasi-stationary state with a finite population in the two-excitation subspace. For equal spacing this is a true stationary state and the excitation remains trapped for all times, while for arbitrary $n_i$ it eventually decays, but on a much longer timescale.

The fast relaxation into a doubly-excited, but non-decaying state is a rather unexpected feature, which is explained by Fig.~\ref{Fig3:Subradiance}(b). Here the possible decay paths are represented in terms of the eigenstates of $H_{\rm eff}$ with different excitation numbers. For equal spacing $n_{i+1}-n_i=x>0$, we find that in the two-excitation manifold there is one exact dark state, satisfying $H_{\rm eff}|D_2\rangle=0$, and additional subradiant states with decay rates $\lesssim 10^{-2} \Gamma_0$. These other subradiant states are almost decoupled from the waveguide due to symmetry. Therefore, they are long-lived, but also hardly populated during the dynamics. In contrast, there is an efficient decay path into state $|D_2\rangle$, which does not decay further. We find that the form of the dark state is,
\begin{equation}
|D_2\rangle = \alpha(x) |egge\rangle - \beta(x) |geeg\rangle,
\end{equation}
where $\alpha(x)/\beta(x)=f_{K_0}(x)/f_{K_0}(3x)\geq1$~\cite{supp}. This state is not invariant under the inversion, $|g\rangle\leftrightarrow |e\rangle$, which explains, why it is possible to have different rates for decaying into and out of it. Similar states also exist for a larger number of emitters, emerging from the combination of long-range interference and the presence of additional correlations, $\sim f_{K_0}(n_i,n_j)$, which vary considerably across the ensemble.

\begin{figure}[t]
    \begin{center}
        \includegraphics[width=\columnwidth]{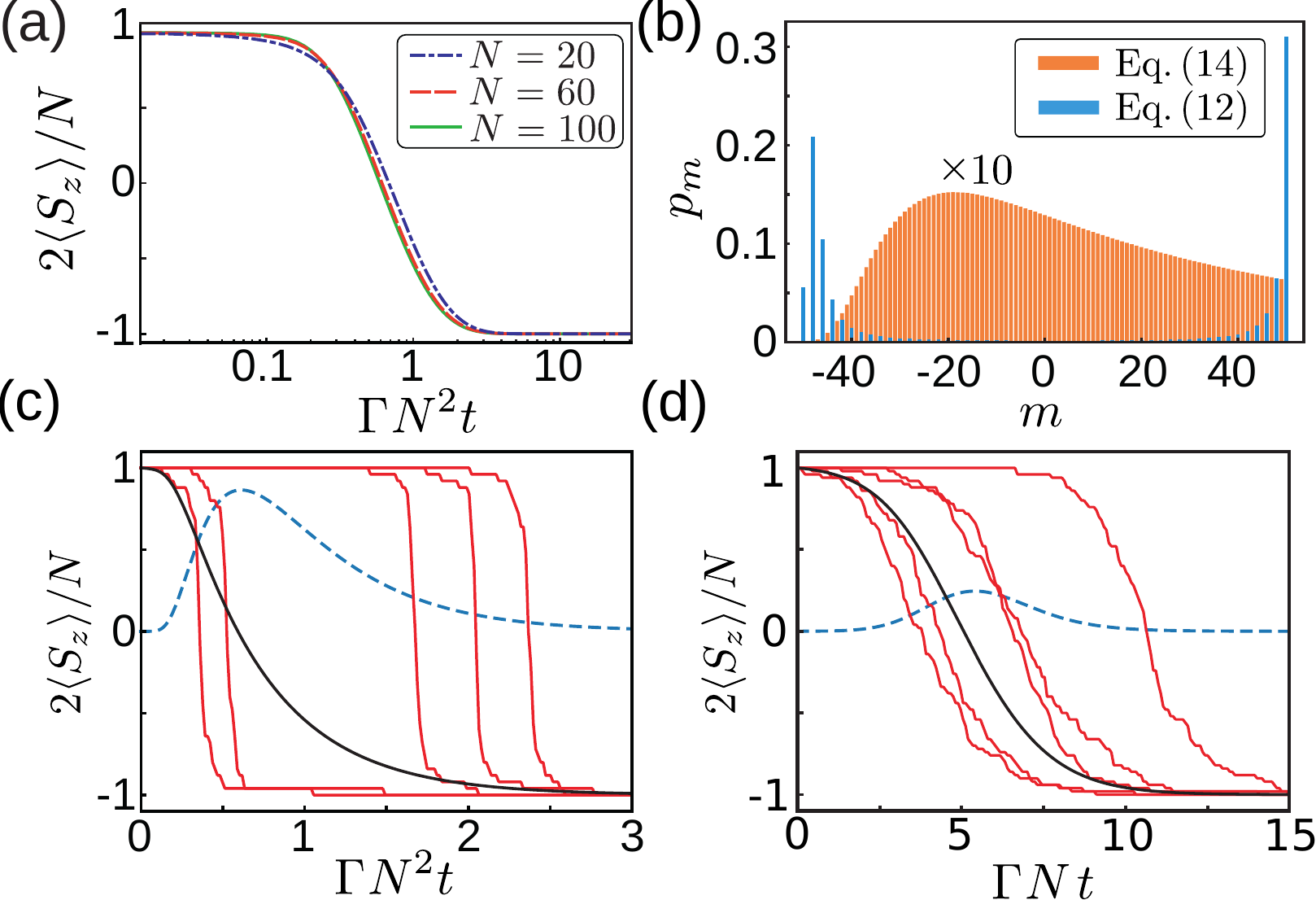}
        \caption{(a) The evolution of $\langle S_z\rangle$ predicted by Eq.~\eqref{master} for an initially fully excited ensemble of TLSs.
        (b) Snapshot of the populations $p_m$ for a super-correlated (blue) and regular superradiant (orange) decay, evaluated when $\langle S_z\rangle  =0$. (c) and (d) Example trajectories as obtained from a stochastic simulation of Eq.~\eqref{master} and Eq.~\eqref{eq:master_SR}. The thick lines show the corresponding average values of $2\langle S_z\rangle/N$ (black) and $4(\langle S_z^2\rangle - \langle S_z\rangle^2)/N^2$ (dashed blue). In (b)--(d) we have used $N=100$.}
        \label{Fig4:Superradiance}
    \end{center}
\end{figure}

\emph{Collective-spin limit.}---From the results of Fig.~\ref{Fig3:Subradiance}(a), we already see that the rate of emission is enhanced when the spacing between emitters is small. Therefore, we consider next the special case where all TLSs are located in the same lattice site and collective effects are most pronounced. In this limit the ME reduces to
\begin{equation}
\dot \rho=\frac{\Gamma}{2}\left(2S_{-}^{2}\rho S_{+}^{2}-\rho S_{+}^{2}S_{-}^{2}-S_{+}^{2}S_{-}^{2}\rho\right),
\label{master}
\end{equation}
where $\Gamma=\Gamma_0 f^2_{K_0}(0)$ and $S_-=\sum_i \sigma_i^-$ is the collective spin lowering operator. Since ME~\eqref{master} conserves the total spin, we can label all the states involved in the dynamics by their spin projection quantum number, $S_z|m\rangle =m |m\rangle$, where $|m=N/2\rangle=|e_1\dots e_N\rangle$ is the fully excited initial state. This leads to a reduced equation for the populations $p_m=\langle m|\rho|m\rangle$,
\begin{equation}
\dot p_m = -\Gamma_{m,m-2} p_m + \Gamma_{m+2,m} p_{m+2},
\end{equation}
where $\Gamma_{m,m-2}= \Gamma |\langle m-2| S^2_-|m\rangle|^2$.
In Fig.~\ref{Fig4:Superradiance}(a), we use this equation to evaluate the collective decay of a large ensemble of TLSs. The non-exponential and accelerated decay is reminiscent of regular Dicke superradiance described by the ME~\cite{Gross1982main,SHbook}
\begin{equation}
\dot \rho=\frac{\Gamma}{2}\left(2S_{-}\rho S_{+}-\rho S_{+}S_{-}-S_{+}S_{-}\rho\right),
\label{eq:master_SR}
\end{equation}
but there are  important qualitative differences. Firstly, at short times, where $m\approx N/2$, the decay rate scales as $\Gamma_{m,m-2}\sim N^2$. This is $N$ times faster than for $N$ independent TLSs and shows that even in the initial stage of the evolution, the dynamics is dominated by correlations. For states near the equator of the Bloch sphere, $m\approx 0$, the rates then scale as  $\Gamma_{m,m-2}\sim N^4$ compared to the $N^2$ scaling for regular superradiance. Overall, this results in a strongly reduced decay time of $T_d \sim 1/(\Gamma N^2)$.

More importantly, while the dynamics of Eq.~\eqref{eq:master_SR} can be well-described by a mean-field approximation, $\langle S_z^2\rangle\approx \langle S_z\rangle^2$~\cite{SHbook}, a similar approach for Eq.~\eqref{master} fails to accurately capture the system evolution~\cite{supp}. This can be understood from the snapshot of the populations $p_{m}$ shown in Fig.~\ref{Fig4:Superradiance}(b), which is taken at the half-decay time $T_h$, defined by $\langle S_z\rangle(T_h)=0$. We see that for regular superradiance there is a broad distribution around its mean value. In contrast, the two-photon decay process leads to a bi-modal distribution, where most of the population is in the states with $m\approx \pm N/2$. The intermediate levels are hardly populated, since they decay with a much faster rate. This different behavior can also be clearly seen by looking at individual trajectories of a stochastic ME simulation~\cite{Daley2014}. The red curves in Fig.~\ref{Fig4:Superradiance}(c) and (d) show example trajectories for the two-photon decay process and regular superradiance, respectively.
We see that in the former case,  the time that the system spends near the fully excited state, $T_e\sim 1/N^2$, is considerably longer than the time it takes to transition through all the partially excited states, $T_t\sim 1/N^3$, such that $T_t/T_e\sim 1/N\rightarrow 0$ for large $N$. This means that when measuring the system at random times during the decay, all TLSs are either still found in the excited state or already in the ground state. It is thus more appropriate to speak of \emph{super-correlated} emission rather than just superradiance. This qualitative difference can also be quantified by the correlation parameter
{
\begin{equation}
C = {\max_{t \in [0,\infty)}} \frac{4\left[\langle S_z^2\rangle(t)-\langle S_z\rangle^2(t)\right]}{N^2},
\end{equation}
which measures the maximal correlations between the populations of the TLSs during the decay.} {For independent emitters, $C\approx 0$ in the limit of large $N$, while it reaches a maximal value of $C = 1$ when all the emitters are correlated at the point when $\langle S_z\rangle=0$. We find that $C\approx 0.2$ for superradiance and $C\approx 1-O(1/N)$~\cite{supp} for the super-correlated decay process described by Eq.~\eqref{master}, giving a clear signature of the dynamics discussed above.}

\emph{Discussion and conclusions.}---In summary, we have studied the collective radiance of an ensemble of TLSs coupled to a nonlinear environment. We found that this system supports a strongly correlated decay process outside the scope of conventional super- and subradiance. In the optical domain, implementations of nonlinear photonic lattices have already been proposed for engineering strongly-correlated fluids of light~\cite{Hartmann2006,GreentreeNatPhys2006,AngelakisPRA2007,Carusotto2013,Angelakis2017} and similar ideas can be used to explore these decay processes. Alternatively, superconducting qubits can be coupled to an array of microwave resonators, where embedded Josephson junctions provide a strong nonlinearity~\cite{Leib2012}. In such systems values of $g\lesssim J,U\approx 50$--$200$\,MHz can be achieved with existing technology~\cite{HacohenGourgy2015,Roushan2017,Ma2019}. For $J  = 100$\,MHz the achievable decay rates of around $\Gamma \approx 0.1$--$1$\,MHz~\cite{supp} are still fast compared to the bare qubit decay times of $T_1=10\,\mu$s~\cite{Kjaergaard2019}. The super-correlated limit $N\gg1$ can further be accessed by replacing the qubits by a large ensemble of Rydberg atoms trapped above the resonator array~\cite{Petrosyan2008,Hogan2012,Beck2016,Calajo2017}, {or even above a single nonlinear cavity~\cite{supp} to further reduce the experimental complexity.}

Beyond the specific setting considered in this work, our analysis demonstrates how non-trivial interactions in the environment can strongly modify the qualitative behavior of open quantum systems. In turn, the established relation between collective radiance and few-body effects in the bath can potentially be used as a more general method to probe complex many-body processes through the correlated decay of multiple quantum emitters.

\emph{Acknowledgments.}--- This work is supported by  National Natural Science Foundation of China (Grant No. 11875011); Educational Commission of Jilin Province of China (Grant No.~JJKH20190266KJ); the China Scholarship Council (CSC) Grant No.
201806625045 (ZW);  Austrian Science
Fund (Grant No.~P31701-N27); DK CoQuS {Grant
No.~W1210} (TJ), and an ESQ Fellowship (PK) and Discovery Grant (PR)  of the Austrian
Academy of Sciences (\"OAW).

\onecolumngrid
\clearpage

%
%

\begin{center}
\textbf{\large Supplementary material for: \\
	Super-correlated radiance in nonlinear photonic waveguides}
\end{center}

\setcounter{equation}{0} \setcounter{figure}{0} 
\makeatletter

\renewcommand{\thefigure}{SM\arabic{figure}} \renewcommand{\thesection}{SM%
\arabic{section}} \renewcommand{\theequation}{SM\arabic{equation}}

\section{Two-photon bound states}

In the main text we focus on results obtained for the behavior of a 1D waveguide.
Here, we outline the calculation of the two-photon bound state wavefunctions and
 energies for the general case of a $d$-dimensional tight-binding lattice, showing
  that similar results can be obtained in other lattice geometries. The Hamiltonian
  we consider is given by
\begin{equation}
H_{\rm env}=\sum_{\vec n}\left(
\omega_c a^\dag_{\vec n} a_{\vec n}-\frac{U}{2} a_{\vec n}^\dag
a_{\vec n}^\dag a_{\vec n} a_
{\vec n}\right) - \frac{J}{2} \sum_{\vec n}\sum_{\vec e} \left(a_{\vec n}^\dag
a_{\vec n+\vec e} + a_{\vec n} a_{\vec n+\vec e}^\dag\right),
\end{equation}
where $\vec{n}$ labels the position of a site, the sum over $\vec e$ runs over
all lattice vectors connecting neighboring sites and periodic boundary conditions
 are assumed. Here we will take $\omega_c=0$, which only
 gives a shift of the zero of energy. We consider the general ansatz for
 the 2-photon wavefunction,
\begin{equation}
|\Psi_{\vec K}\rangle= \frac{1}{\sqrt{2}} \sum_{\vec n,\vec m} \Psi_{\vec K}(\vec n,\vec m)
a^\dag_{\vec n} a^\dag_{\vec m} |{\rm vac}\rangle,
\end{equation}
and separate it into center-of-mass and relative coordinates, $\Psi(\vec n,\vec m)
=e^{i\vec K\cdot(\vec n+\vec m)/2}\psi_{\vec K}(\vec n-\vec m)/\sqrt{N_c}$. Here $\vec K$
lies within the first Brillouin zone and for the wavefunction, $\psi_{\vec K}(\vec r )$,
 for the relative coordinate $\vec r=\vec n-\vec m$ we obtain the eigenvalue equation
\begin{equation}\label{eq:Esupp}
E \psi_{\vec K} (\vec r) = - 2 \sum_\alpha J_{\vec K_\alpha}\left[   \psi_{\vec K}
(\vec r+\vec e_\alpha)+\psi_{\vec K} (\vec r-\vec e_\alpha)\right] -
U \delta_{\vec r,0} \psi_{\vec K} (\vec r),
\end{equation}
where  $J_{\vec K_\alpha}= J\cos(\vec K_\alpha /2)$, and the summation
is performed over all of the nearest neighbor sites. The general set of eigenstates can then be obtained from the solution of the corresponding Lippmann-Schwinger equation, as discussed in~\cite{Piil2007,Winkler2004}. Since here we are only interested in the bound states, we solve Eq.~\eqref{eq:Esupp}
by changing to a Fourier representation
\begin{equation}\label{wavefunctiontrans}
\psi_{\vec K}(\vec r)= \frac{1}{\sqrt{N_c}} \sum_{\vec q}
e^{i\vec q\cdot\vec r} \psi_{\vec K}(\vec q),
\end{equation}
and obtain
\begin{equation}\label{wavefunctionq}
\psi_{\vec K}(\vec q)= -  \frac{U}{\sqrt{N_c}} \frac{\psi_{\vec K}
(\vec r=0)}{E+ 4 \sum_{\alpha} J_{\vec K_\alpha} \cos(\vec q_\alpha)}.
\end{equation}
By using $\psi_{\vec K}(\vec r=0)= \frac{1}{\sqrt{N_c}} \sum_{\vec q}
\psi_{\vec K}(\vec q)$ and changing sums into integrals we end up with
the condition
\begin{equation}\label{eq:EigenvaluesGeneral}
\frac{1}{U} = - \frac{1}{(2\pi)^d} \int_{-\pi}^\pi  \,  \frac{ d^d q }{ E + 4 \sum_{\alpha} J_{\vec K_\alpha} \cos(\vec q_\alpha)}.
\end{equation}
It allows us to calculate the properties of the 2-photon bound state for an arbitrary lattice geometry. Below we explicitly calculate the behavior for 1,2 and 3-dimensional square lattices.

\subsection{1D}
In one dimension the integral over $q$ can be solved exactly,
\begin{equation}
\frac{1}{U} =-\frac{1}{2\pi} \int_{-\pi}^\pi  \ \frac{ dq  }
{ E + 4J_K \cos(q)}=  \frac{1}{\sqrt{E^2-16J_K^2}},
\end{equation}
and we obtain the bound-state energy
\begin{equation}
E_{K}^b = - \sqrt{U^2+16 J^2 \cos^2(K/2)}.
\end{equation}
The lowest energy state is therefore the bound state with $K=0$ and
 the band of bound states has a width $\sqrt{U^2+16 J^2}-U$.

{The wave function for $r\neq0$ can be directly obtained from
 Eqs.~\eqref{wavefunctiontrans}-\eqref{wavefunctionq} as
\begin{eqnarray}
\psi_{K}(r)&=& \frac{1}{\sqrt{N_c}}\sum_{q}e^{iqr}\psi_{K}(q)=-\frac{U \psi_{K}(0)}{2\pi} \int_{-\pi}^{\pi}dq\frac{e^{iqr}}{E+4J_K\cos(q)}.
\label{rfunction}
\end{eqnarray}
Performing the integration, we obtain the normalized wave function as
\begin{equation}
\psi_{K}(r)=\sqrt{\tanh(\lambda_K^{-1})}e^{-\frac{|r|}{\lambda_K}},
\end{equation}
where the width of the bound state is given by $\lambda_K^{-1}=\mathrm{asinh}(\mathcal{U}_k)$, with $\mathcal{U}_k=U/4J_K$.

\begin{figure}
\begin{centering}
\includegraphics[width=0.9\columnwidth]{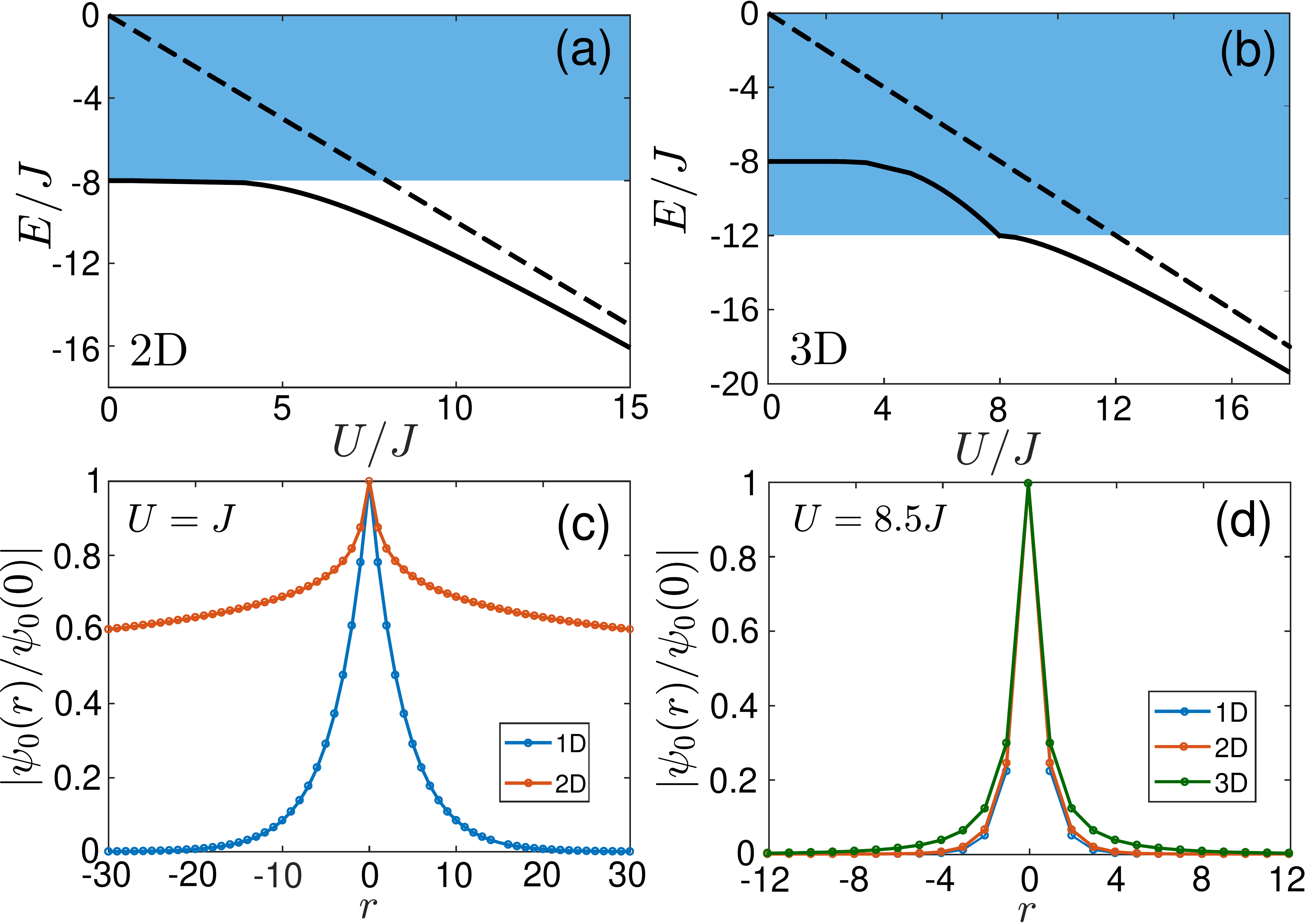}
\end{centering}
\caption{(a) and (b): The boundaries of the energy band for two-photon bound
states for 2D and 3D. In the shaded blue region, scattering states also exist. (c) and (d): The wavefunction of the bound state for
small and large $U$. Note that for $ E < -8J $ no bound state exists for small $ U $ in 3D. }
\label{bound}
\end{figure}

\subsection{2D and 3D}

In higher dimensions the integral in Eq.~\eqref{eq:EigenvaluesGeneral} can
 no longer be solved analytically. However, for small $U$ the bound states
have an energy only slightly below the propagation band and hence the main
 contribution to the integral will come from $|\vec q| \ll 1$.  This then allows us, for
 $\vec K=0$, to approximate
\begin{equation}
\frac{1}{U} \simeq - \frac{1}{(2\pi)^d} \int_{-\pi}^\pi \frac{ d^d q }
{ (E+4Jd)-  2J  |\vec q|^2}.
\end{equation}
In the 2D case we find
\begin{equation}
\frac{1}{U} \simeq  -\frac{1}{2\pi} \int_{0}^\pi dq \,  \frac{q}{ E+8J-2Jq^2}=
 \frac{1}{8J\pi}  \ln\left( 1- \frac{2J \pi^2}{|E+8J|}\right).
\end{equation}
Therefore, while a bound state exists for arbitrary small $U$, its energy
\begin{equation}
E_{K=0}^b\simeq -8J - 2J \pi^2 e^{-\frac{8J\pi}{U}}
\end{equation}
is exponentially close to the band edge. In contrast for a 3D lattice the equivalent expression is
\begin{equation}\label{eq:OneOverU}
\frac{1}{U} \simeq  \frac{1}{2\pi^2} \int_{0}^\pi  dq \,
\frac{  q^2 }{E+12J-2J  q^2}\approx   \frac{1}{4\pi J}
\left[ 1-\sqrt{\frac{|E+12J|}{2J}}\right] .
\end{equation}
In this case, a bound state only exists for couplings above a critical value of the nonlinearity $U\gtrsim 4\pi J$.  Note that in 3D the integral in Eq.~\eqref{eq:OneOverU} depends substantially on $q$ values away from the band edge, where the quadratic approximation of the dispersion relation is no longer valid. Therefore, the predicted value for this critical interaction strength is not very accurate.

In Fig.~\ref{bound}(a) and (b), we show the exact boundaries of the two-photon
 bound state band as a function of $U/J$ for both 2D and 3D. The upper
  bound is given by $E=-U$, independent of the lattice, and the lower
  bound is given by the numerical solution of Eq.~\eqref{eq:EigenvaluesGeneral}, as discussed above.
}{Furthermore, we can also numerically obtain the wavefuction from
Eqs.~\eqref{wavefunctiontrans}-\eqref{wavefunctionq}. This is shown as a
function of the relative coordinate $r$ for both small and large $U$ in Fig.~\ref{bound}(c) and
 (d), respectively. Since the problem has rotational symmetry we only show the
  profile along the $x$ direction, that is, $r=r_x$. The size of the bound
  state decreases with increasing $U$. At small $U$ (where the bound state does not exist in 3D) the wavefunction extends over more
   lattice sites in 2D than in 1D, while at large $U$ the form of the
   wavefunction does not significantly depend on the lattice dimension.}

{
\section{Single-emitter decay}\label{sec:SingleEmitterDecay}
Before we return to the correlated decay of multiple emitters, we briefly discuss the decay of a single emitter with a frequency below the propagation band, i.e., $\delta_{0}=\omega_c-\omega_e -2J>0$. By assuming that the photon loss rate $\kappa$ is smaller than all the other frequency scales, $\kappa\ll \delta_0, g, J$, the rate of decay for a single atom is given by
\begin{equation}\label{eq:Gamma1}
\Gamma_1 \simeq T_1^{-1} + \sin^2(\theta) \kappa,
\end{equation}
where $T_1$ is the bare decay time of the emitter. The second part arises from the hybridization between the emitter and the photonic states in the waveguide, where $\theta$ is the mixing angle (see, for example, Ref.~\cite{GC} and references therein).
The photonic fraction is given by~\cite{GC}
\begin{equation}
\sin^2(\theta)= \frac{g^2}{g^2+(E-\omega_c)^2(1-4J^2/(E-\omega_c)^2)^{3/2}},
\end{equation}
where $E<\omega_c-2J$ is the energy of the atom-photon bound state and is given by the solution of
\begin{equation}
E-\omega_c+2J+\delta_0 = \frac{g^2}{\sqrt{(E-\omega_c)^2-4J^2}}.
\end{equation}
For $g\ll |\delta_0|$ the energy $E$ coincides essentially with the energy of the emitter, $E\simeq \omega_c-2J-\delta_0$. For $\delta_0\gg 2J$ we then obtain the expected result for a far detuned emitter
\begin{equation}
\sin^2(\theta)\approx   \frac{g^2}{\delta_0^2}.
\end{equation}
In the more relevant limit $\delta_0\ll 2J$ we obtain instead
\begin{equation}\label{eq:AppSin2}
\sin^2(\theta)\approx   \frac{g^2}{4\sqrt{J\delta_0^3}}.
\end{equation}
Based on this result, we obtain a photonic fraction of $\sin^2(\theta)\approx 0.0125$ for the values of $g=0.02J$ and $\delta_0=0.04J$ assumed in Fig. 1(c) in the main text. For a value of $\kappa/J=3\times 10^{-4}$ the estimated single atom decay $\Gamma_1/J\approx 3.75 \times 10^{-6}$ is in a reasonable agreement with the decay observed in the full numerical calculations.

\begin{figure}
\begin{centering}
\includegraphics[width=0.9\columnwidth]{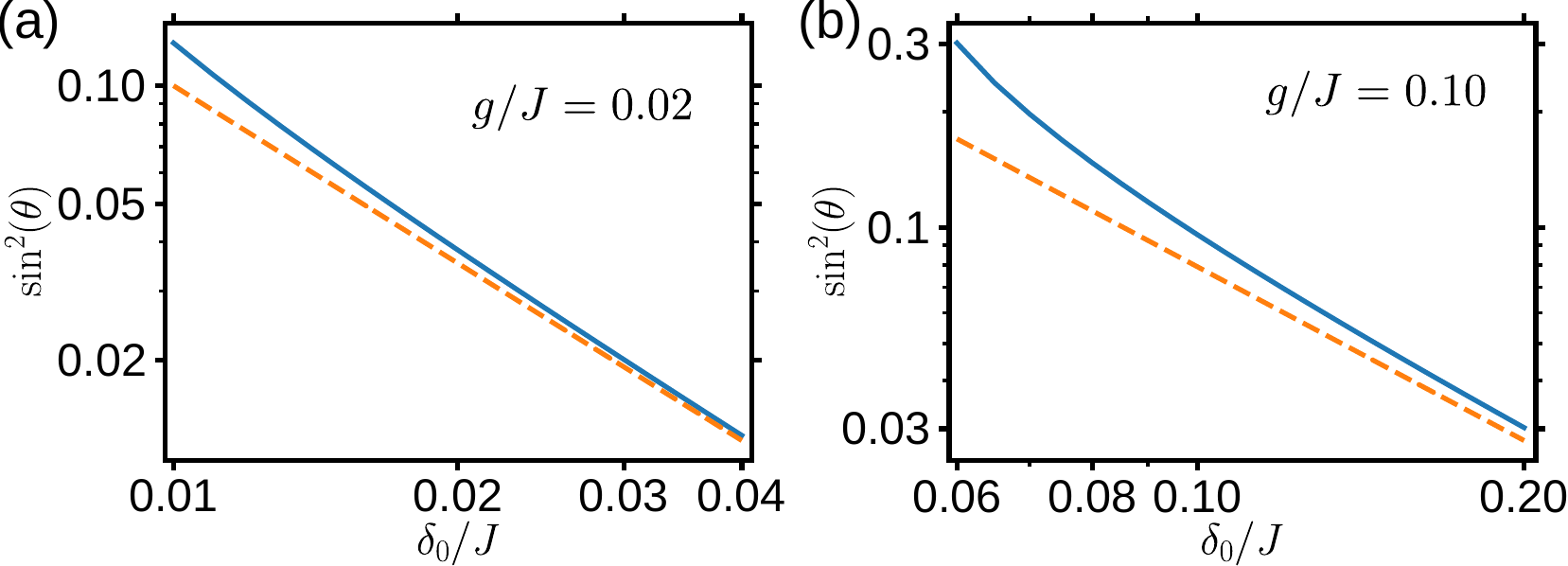}
\end{centering}
\caption{{Plot of the photon component $\sin^2(\theta)$ of an atom-photon bound state~\cite{GC}, which is used to evaluate the residual single-emitter decay rate in Eq.~\eqref{eq:Gamma1}. For the two plots, the values of (a) $g/J=0.02$ and (b) $g/J=0.1$ are assumed. These values are also used in Fig. 1(c) in the main text. The blue solid line shows the exact result for $\sin^2(\theta)$ and the orange dashed line the approximate analytic scaling given in Eq.~\eqref{eq:AppSin2}.} }
\label{fig:PhotonComponent}
\end{figure}

For $g\sim \delta_0$ a simple analytic expression for the photonic fraction is no longer possible. Therefore, in Fig.~\ref{fig:PhotonComponent} we plot the dependence of $\sin^2(\theta)$ on the detuning $\delta_0$ for $g/J=0.02$ and $g/J=0.1$. We see that as soon as $\delta_0\gtrsim g$ the photonic fraction and, consequently, the single emitter decay rate are strongly reduced.
}

\section{Wigner-Weisskopff approximation for two emitters}

In this section we show how to obtain the simplified model of the decay of two
emitters presented in the main text.  Both emitters are initialized in the
excited state $|e\rangle$ and the waveguide in the vacuum state $|\text{vac}\rangle$.
Since the Hamiltonian conserves the total excitation number we can write the
wavefunction of the whole system as
\begin{equation}
\begin{split}
|\psi_2(t)\rangle= & e^{-2i\omega_{e}t}\Big[c_e(t)\sigma_1^+\sigma_2^+ +
\sum_{K}c_{K}(t)B_K^\dag + \sum_k ( c_{1k}(t) \sigma_1^+ + c_{2k}(t)
\sigma_2^+) a^\dag_k  \Big] |g,g,{\rm vac}\rangle + |\psi_2^{(s)}(t)\rangle.
\end{split}
\end{equation}
Here $c_e(t)$ is the amplitude of the state with both emitters excited,
$c_{K}(t)$ is the amplitude of the two-photon bound state with wavevector $K$,
and the $c_{ik}(t)$ are the amplitudes of state with emitter $i$ excited and a single photon state
with wavevector $k$ in the waveguide.  The last term, $|\psi_2^{(s)}(t)\rangle$,
 accounts for the part of the wavefunction with support in the subspace of
 two-photon scattering states. Since these states are off-resonant and only
 excited through high-order processes, we omit this component in the following
 analysis. Based on a comparison with full numerical simulations, we find that
  this approximation is well justified in the parameter regimes of interest.
  For the other amplitudes we obtain the set of coupled equations
\begin{eqnarray}
i\dot{c}_e(t) & = & \frac{g}{\sqrt{N_c}}  \sum_{k} \left[  e^{ikn_{2}} c_{1k}(t)
 +e^{ikn_{1}} c_{2k}(t)\right],\label{ce}\\
i\dot c_{1k}(t)&=& \delta_k c_{1k}(t) +  \frac{g}{\sqrt{N_c}} e^{-ikn_{2}} c_e(t)
+ g\sum_K  M(k,n_1,K) c_K(t),\label{c1k}\\
i\dot c_{2k}(t)&=& \delta_k c_{2k}(t) +  \frac{g}{\sqrt{N_c}} e^{-ikn_{1}} c_e(t)
+ g\sum_K  M(k,n_2,K) c_K(t),\label{c2k}\\
i\dot c_K(t)&=& \Delta_K c_{K}(t) + g\sum_k  \left[ M^*(k,n_1,K) c_{1k}(t)
+M^*(k,n_2,K) c_{2k}(t)\right]\label{cK}.
\end{eqnarray}
Here we have introduced the single-photon and two-photon detunings,
$\delta_k =\omega_k-\omega_e$ and $\Delta_K=E_K^b-2\omega_e$, and the
coupling matrix element
\begin{equation}
M(k,n,K)= \langle {\rm vac}| a_k a_n B^\dag_K|{\rm vac}\rangle =
 \frac{\sqrt{2}}{N_c}\sum_m e^{-ikm} e^{iK(n+m)/2}\psi_K^b(n-m),
\end{equation}
note here that $a_k^{\dagger}$ creates a photon with wavevector $k$ while $a_n^{\dagger}$ creates a photon localized at site $n$. We are interested in the regime where single photon processes are
suppressed, $\delta_k\gg |\Delta_K|,g$. This allows us to adiabatically
eliminate the amplitudes $c_{ik}$ by setting $\dot c_{ik}=0$. Therefore,
\begin{eqnarray}
c_{1k} &\simeq &-\frac{g}{\delta_k} \left[ \frac{1}{\sqrt{N_c} }
 e^{-i k n_2} c_e(t) + \sum_K M(k,n_1,K) c_K(t)  \right],\\
c_{2k}&\simeq &-\frac{g}{\delta_k}\left[ \frac{1}{\sqrt{N_c}}
e^{-ikn_1} c_e(t) + \sum_K M(k,n_2,K) c_K(t)  \right].
\end{eqnarray}
{Note that this adiabatic elimination is only valid for $|\Delta_K|\ll \delta_k$, i.e., for near-resonant two-photon states below the propagation band. However, these are exactly the modes that play the dominant role in the decay dynamics.}

After reinserting these expressions back into the equations of
motion for $c_e(t)$ and $c_K(t)$, we obtain a pair of coupled equations
for the amplitudes of interest
\begin{eqnarray}
i\dot{c}_e(t)& =& 2\delta \omega_{e}  c_e(t)-\frac{g^2}{J\sqrt{N _c}}
\sum_{K}   e^{i K(n_1+n_2)/2} f_K(n_1,n_2) c_K(t),\\
i\dot{c}_K(t) &=&\sum_{K'}  \left[  \Delta_K \delta_{K,K'} + G(K,K')
\right]c_{K'}(t)- \frac{g^2}{J\sqrt{N _c}}  e^{-i K(n_1+n_2)/2} f_K(n_1,n_2) c_e(t).
\end{eqnarray}
We see that the virtual coupling to the one-photon states introduces
 a Stark-shift to the frequency of the doubly excited state,
\begin{equation}
\delta \omega_{e}=  -\frac{1}{N_c}\sum_k \frac{g^2}{\delta_k},
\end{equation}
 and frequency shifts and cross couplings between the two-photon states,
  described by the diagonal and off-diagonal elements of the matrix
\begin{equation}
 G(K,K') =-  \sum_k\sum_{i=1,2} \frac{g^2}{\delta_k} M^*(k,n_i,K) M(k,n_i,K').
\end{equation}
An effective coupling is also introduced between the two emitters and the two-photon states
 with a normalized matrix element
\begin{equation}
f_K(n_1,n_2) =\sum_{k} \frac{ J }{\delta_k} \left[e^{-ikn_{2}}
M^*(k,n_1,K)+e^{-ikn_{1}}M^*(k,n_2,K)\right] e^{i K(n_1+n_2)/2},
\label{ffunction}
\end{equation}
which depends on the distance $r=n_1-n_2$ only.
By using the definition $M^*(k,n,K)= \langle {\rm vac}|  B_K a^\dag_n a^\dag_k|{\rm vac}\rangle$ and rewriting the sum over $k$ as
\begin{equation}
i \sum_k \frac{e^{-ikn_{2}}}{\delta_k} a^\dag_k = \lim_{\epsilon\rightarrow 0} \int_0^\infty  d\tau   \sum_k e^{-ikn_{2}} a^\dag_k e^{i\delta_k\tau} e^{-\epsilon \tau} = \sqrt{N_c} \int_0^\infty  d\tau   a^\dag_{n_2}(\tau) e^{-i\omega_e \tau},
\end{equation}
where $a^\dag_n(t)$ is the free evolution of the single-photon creation
operator, we can express this amplitude as
\begin{equation}
f_K(n_1,n_2)= -i \sqrt{N_c} J \int_0^\infty d\tau \, e^{i K(n_1+n_2)/2}  e^{-i\omega_e \tau} \langle {\rm vac}| B_K(  a^\dag_{n_2} a^\dag_{n_1}(\tau)  +  a^\dag_{n_1}a^\dag_{n_2}(\tau))|{\rm vac}\rangle,
\end{equation}
as given in the main text. {Eq.~\eqref{ffunction} shows that $f_K(n_1,n_2)$
is a real number for any $K$, $n_1$ and $n_2$.} {In terms of the wave function of two-photon bound states, the element $f_K(n_1,n_2)$ can be expressed as
\begin{equation}
f_{K}(n_{1}, n_{2})=\dfrac{2\sqrt{2}}{N_c}\sum_{k}\frac{J\cos[(k-K/2)(n_{1}-n_{2})]L_{k,K}}{\delta_{k}},
\end{equation}
where
\begin{equation}
L_{k,K} = \sum_m e^{-i(k-K/2)m}\Phi_K(m) = \dfrac{\sinh(\lambda_{K}^{-1})}{\cosh(\lambda_{K}^{-1}) - \cos(k-K/2)}\Phi_{K}(0).
\end{equation}}

\begin{figure}
\begin{centering}
\includegraphics[width=0.9\columnwidth]{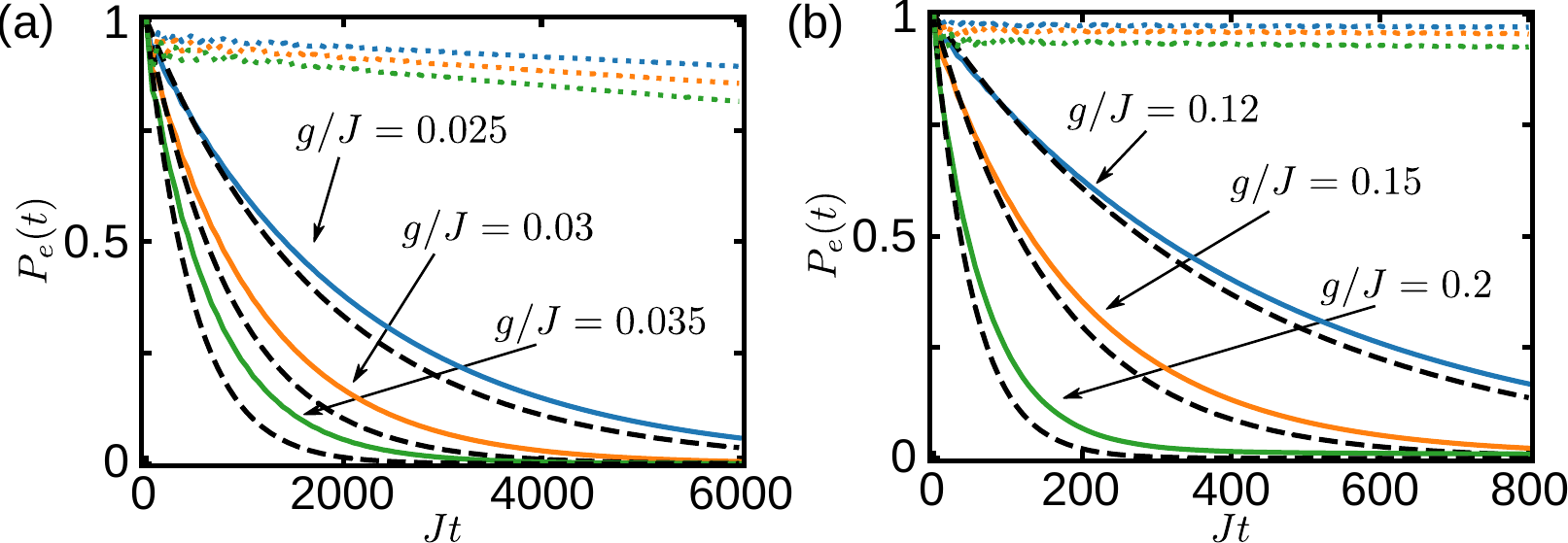}
\end{centering}
\caption{Evolution of the excited state population $P_e(t)$ for
$N = 1$ (dotted) and $N = 2$ (solid) emitters with frequency
$\omega_e < \omega_c-2J$. The parameters are set as $\kappa/J=3\times10^{-4}$
and (a) $U/J=1$, $(\omega_e-\omega_c)/J=-2.04$, $K_0\approx0.1\pi$ and
(b) $U/J=4$, $(\omega_e-\omega_c)/J=-2.45$, $K_0\approx0.5\pi$. The values of $g/J$ are as indicated and we have set $n_1=n_2$.}
\label{dynamics}
\end{figure}

To proceed, we reabsorb the Stark shift $\delta \omega_e$ into the definition
of the emitter frequency and neglect the couplings $G(K,K')$ between the
two-photon amplitudes. The validity of this approximation is again
verified by a comparison with full numerical simulations. Under this
approximation we obtain
\begin{equation}
c_K(t) \simeq   i\frac{g^2}{J\sqrt{N _c}}  e^{-i K(n_1+n_2)/2}
f_K(n_1,n_2)\int_0^t e^{-i\Delta_K (t-t')} c_e(t')dt'
\end{equation}
and
\begin{equation}\label{eq:SuppCedot}
\dot{c}_e(t)= -   \frac{g^4}{J^2} \int_0^t dt' \left[  \frac{1}{N _c}
\sum_{K}  f^2_K(n_1,n_2) e^{-i\Delta_K (t-t')}\right] c_e(t').
\end{equation}
In a final step, we make the usual Markov approximation by assuming
that $f_K(n_1,n_2)$ is a slowly varying function around the resonant
wavevector $K_0$ and replacing the remaining expressions in the square
brackets by a $\delta$-function, i.e.,
\begin{equation}\label{eq:SuppCedotMarkov}
\frac{1}{N _c}  \sum_{K}  f^2_K(n_1,n_2) e^{-i\Delta_K (t-t')}
\approx  \frac{1}{J} \tilde \rho(K_0) \delta (t-t')f^2_{K_0}(n_1,n_2).
\end{equation}
Here we have introduced the normalized density of two-photon
bound states
\begin{equation}
\tilde \rho(K) = \frac{J}{v_g(K)} =
\frac{\sqrt{U^2+16J^2\cos^2(K/2)}}{4J\sin(K)},
\end{equation}
{where
\begin{equation}
v_g(K) = \frac{\partial E_K^b}{\partial K} = \frac{4J^2\sin(K) }{\sqrt{U^2+16J^2\cos^2(K/2)}},
\end{equation}
is the group velocity of the emitted photon pairs.} Altogether we end up with
\begin{equation}
\dot c_e(t)= -\frac{\Gamma}{2}c_e(t),\qquad \Gamma=
\frac{2g^4}{J^3} \tilde \rho(K_0) f^2_{K_0}(n_1,n_2).
\end{equation}
The final approximation which we make is that $P_e(t)\simeq|c_e(t)|^2$, i.e.\ we ignore the contributions of the single excitation sector, which we again numerically find to be small. We therefore show that under the approximations described above the two-excitation
probability decays exponentially with a rate determined by $\Gamma$.

{
\subsection{Validity of the Markov approximation}\label{sec:SuppMarkovApprox}
In going from Eq.~\eqref{eq:SuppCedot} to Eq.~\eqref{eq:SuppCedotMarkov} we have made a Markov approximation by replacing the sum over $K$ by a $\delta$-function in time. In physical terms, this approximation is valid when the emitted photon pairs leave the interaction region on a timescale $\tau_c$ that is much shorter than the inverse of the effective coupling rate $g_{\rm eff}$ between the emitters and the photon pairs. By assuming $f_{K_0}\sim O(1)$ and for small separations $n_1-n_2$ we can set $g_{\rm eff}\approx g^2/J$ and estimate the correlation time by the inverse of the group velocity, $\tau_c\approx 1/v_g(K_0)$.
Therefore, we obtain the following condition for the validity of the Markov approximation
\begin{equation}\label{eq:MarkovCondition}
\frac{g^2}{J^2} \ll  \frac{4J\sin(K_0)}{\sqrt{U^2+16J^2\cos^2(K_0/2)}}.
\end{equation}
We can compare this result with the condition required for a Markovian decay of single emitters in a regular waveguide (see, for example, Appendix A of Ref.~\cite{GC}), which reads
\begin{equation}
\frac{g}{J} \ll  2\sin(k_0).
\end{equation}
Here, $k_0$ is the resonant wavevector for single-photon emission. Therefore, we find that for not too large values of $U/J$ the conditions for the validity of the Markov approximation for single-photon emission and two-photon emission are very similar. This is because in the latter case both the effective coupling and the group velocity are reduced. In all the examples discussed in this work, Eq.~\eqref{eq:MarkovCondition} is satisfied.}

{In the derivation of the master equation for $N>2$ emitters given below, the total size $\Delta n=|n_N-n_1|$ of the ensemble might no longer be negligible. In this case an additional factor $1/\Delta n$ must be added to the right-hand side of Eq.~\eqref{eq:MarkovCondition}, which accounts for the fact that the photons need longer to pass through the whole system~\cite{GC}.}

\subsection{Numerical verification of the Wigner-Weisskopff approximation for stronger couplings}
In the main text, we only show the dynamics of $P_e(t)$ for relatively weak emitter-waveguide couplings, $g$;  here, in Fig.~\ref{dynamics}, we demonstrate that the expression above also describes well  the behavior at larger values of $g$. The solid lines show the full numerical results and the dashed lines the exponential decay with rate $\Gamma$. {For the full numerical result, we simulated Eqs.~\eqref{ce}--\eqref{cK} in momentum space for a photonic lattice with $N_c=3001$ sites and periodic boundary conditions.  For each cavity we have also included a small loss rate $\kappa$ with $\kappa/J=3\times10^{-4}$.}
This comparison shows that even for stronger couplings between the emitters and the waveguide, the behavior of the decay of the population is accurately predicted. {Also, the observed decay of the single emitter states is in good agreement with the estimates for $\Gamma_1$ discussed in Sec.~\ref{sec:SingleEmitterDecay}.}

\begin{figure}
\begin{centering}
\includegraphics[width=0.5\columnwidth]{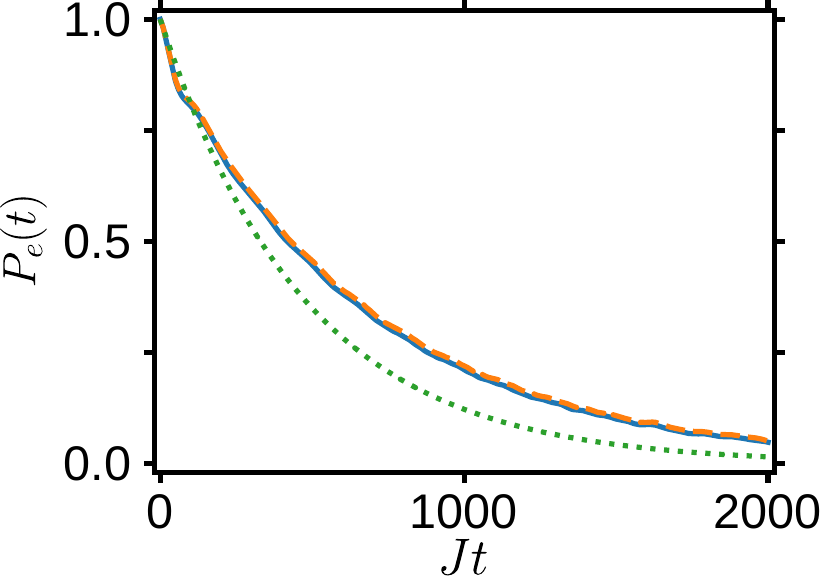}
\end{centering}
\caption{{Evolution of the excited state population $P_e(t)$ for the case with (blue solid line) and without (orange dashed line) taking two-photon scattering state into consideration. The green dotted line is the analytic prediction. The parameters are set as $\kappa/J=3\times10^{-4}$,
 $U/J=1$, $(\omega_e-\omega_c)/J=-2.04$, $ g/J = 0.035 $, $K_0\approx0.1\pi$, $n_1=n_2$. The size of the waveguide is $N_c=601$.}}
\label{dynamicscompare}
\end{figure}

{
\subsection{Validity of neglecting two-photon scattering states}
In the above analysis we have neglected the two-photon scattering states, since they have energies $E_{\rm scat}>2\omega_c-4J$ and for emitter frequencies $\omega_e<\omega_c-2J$ these states are never resonantly excited. We also emphasize that these states have been omitted in the `exact' simulations in Fig. 1(c) in the main text and in Fig.~\ref{dynamics} above. This allowed us to simulate larger arrays and longer times.

To justify this approximation we perform additional simulations for a smaller system of size $N_c=601$ in the position basis. In this case the influence of the scattering states is fully taken into account. In Fig.~\ref{dynamicscompare} we show the results for the decay dynamics obtained in real space and without the inclusion of the two-photon scattering states, Eqs.~\eqref{ce}--\eqref{cK}. While for these parameters one observes small quantitative differences between the exact dynamics and the analytic predictions from Wigner-Weisskopff theory, these differences are not related to the two-photon scattering states. Indeed, neglecting or including the scattering states has no observable influence on the dynamics. Note that in this example $U/J=1$, where the emitters are already tuned quite close to the propagation band. We have verified that for any larger $U$ the influence of the scattering states is even smaller.

In summary, we conclude that neglecting the two-photon scattering states in the analytic derivations and numerical simulations is a well-justified approximation as long as emitter frequencies outside the propagation band are considered.

}

\section{Master equation}
In this section, we outline the derivation of the master equation,
Eq.~(8) in the main text. For this we consider the same
weak-coupling conditions as above, where the off-resonant single photon
states can be adiabatically eliminated. For multiple atoms it is convenient
to express the effective equations of motion for the state amplitudes
in terms of an effective interaction Hamiltonian $H_{\rm int}$ between
the TLSs and the two-photon bound states. For multiple emitters at
locations $n_i$, this effective interaction Hamiltonian is given by
\begin{equation}
H_{\rm int} \simeq i \frac{g^2}{J} \sum_{i,j=1}^N \left[ \sigma_i^-
\sigma_j^- \mathcal{B}^\dag_{n_i,n_j} -  \sigma_i^+\sigma_j^+ \mathcal{B}_{n_i,n_j}\right],
\end{equation}
where, as above, the scattering terms $G(K,K')$ have been neglected.
Here we have introduced the creation operator
\begin{equation}
\mathcal{B}^\dag_{n_i,n_j}= \frac{1}{\sqrt{N_c}} \sum_K e^{-iK(n_i+n_j)/2} f_K(n_i,n_j) B_K^\dag ,
\end{equation}
which creates the two-photon wavepacket coupled to two emitters
located at positions $n_i$ and $n_j$. We emphasize that this Hamiltonian
is only valid when there is, at most, one bound photon pair within the
interaction region. However, this condition is fully consistent with
the following Markovian treatment of the system-bath coupling, where
it is assumed that an emitted photon pair leaves the interaction region
before the system has time to evolve. Note that, for $U\sim J$ and
away from the band edges, the group velocity of the two-photon bound
states, $v_g(K)= \partial E_K^b/\partial K\sim J $, is similar to that
of single photons in the waveguide, while the two-photon emission rate
is considerably smaller. Therefore, the validity of the Markovian master
equation for the two-photon decay is not more stringent than that of a
master equation treatment of regular waveguide QED. {See Ref.~\cite{GC} and the discussion given in Sec.~\ref{sec:SuppMarkovApprox} above.}

Using $H_{\rm int}$ as the relevant system-bath interaction Hamiltonian,
we follow the standard procedure and derive a master equation for the density
operator $\rho$ of the TLSs only~\cite{HB},
 \begin{equation}
 \dot{\rho}=-\int_{0}^{\infty}d\tau \, {\rm Tr}_{c}\{[H_{\rm int}(t),
 [H_{\rm int}(t-\tau),|\rm vac\rangle\langle {\rm vac}| \otimes\rho(t)]]\}.
 \end{equation}
After evaluating the trace over the photon degrees of freedom, the
master equation can be written in the form
 \begin{equation}
 \dot{\rho}=\sum_{i,j,k,l}A_{ij,kl}(\sigma_{i}^{-}
 \sigma_{j}^{-}\rho\sigma_{k}^{\dagger}\sigma_{l}^{\dagger}
 -\rho\sigma_{k}^{\dagger}\sigma_{l}^{\dagger}\sigma_{i}^{-}\sigma_{j}^{-})
 +A_{ij,kl}^{*}(\sigma_{i}^{-}\sigma_{j}^{-}\rho\sigma_{k}^{\dagger}
 \sigma_{l}^{\dagger}-
 \sigma_{k}^{\dagger}\sigma_{l}^{\dagger}\sigma_{i}^{-}\sigma_{j}^{-}\rho),
 \end{equation}
 where the relevant correlation function is given by
 \begin{eqnarray}
 A_{ij,kl}&=&\frac{2g^{4}}{J^2}\int_{0}^{\infty}d\tau \, \langle{\rm vac}|
 \mathcal{B}_{n_i,n_j} (t) \mathcal{B}^\dag_{n_k,n_l}(t-\tau)|{\rm vac}\rangle e^{2i\omega_{e}\tau}\nonumber\\&=&\frac{2g^4}{J^3} \tilde \rho(K_0)f^*_{K_{0}}(n_{i}-n_{j})f_{K_{0}}(n_{k}-n_{l})e^{i K_{0}
 \left|(n_{k}+n_{l})-(n_{i}+n_{j})\right|/2}.
 \end{eqnarray}
 Finally, we can regroup the individual terms into the form given in
 Eq.~(8) in the main text.

\section{Dephasing and single-emitter decay}
\begin{figure}
\begin{centering}
\includegraphics[width=0.5\columnwidth]{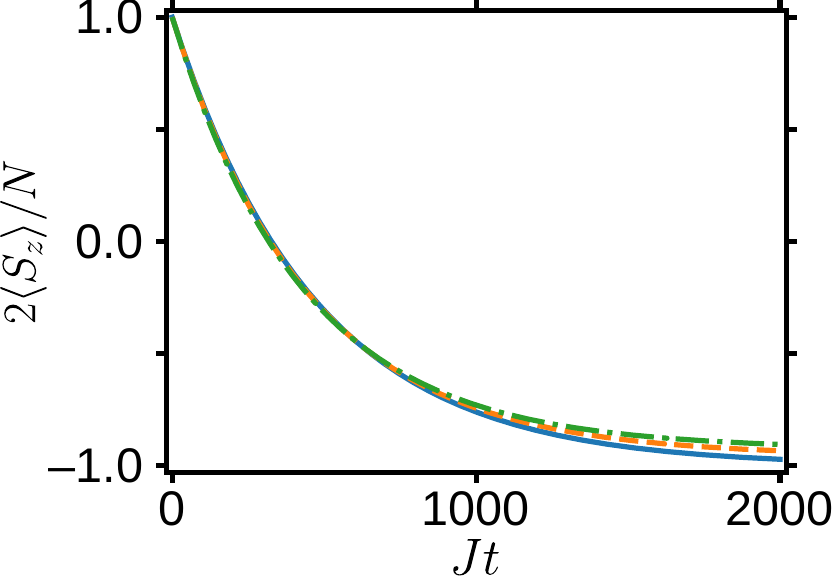}
\end{centering}
\caption{{Dynamics of two-atom system based on the master equation with only 2-photon decay (blue solid line) and with 2-photon decay, single-atom decay and dephasing. In the latter case we have assumed $ T_1^{-1} = 0.1\Gamma $ and $ T_2^{-1} = 0.2\Gamma $ (orange dashed line) and $ T_1^{-1} = 0.2\Gamma $ and $ T_2^{-1} = 0.4\Gamma $ (green dashed-dotted line). The other parameters are set as $U/J=1$, $(\omega_e-\omega_c)/J=-2.04$, $K_0\approx0.1\pi$, $n_1=n_2$, $g=0.035J$. For these parameters we obtain $\Gamma/J\approx 5\times 10^{-4}$ such that for $ J/(2\pi) \approx 100\,\mathrm{MHz} $, the decay times in the second example correspond to  $ T_{1} \approx 15\,\mu\mathrm{s} $ and $ T_2 \approx 7.5\,\mu\mathrm{s} $.}
}
\label{masterequation}
\end{figure}

{In all our numerical simulations we have neglected the effect of the bare decay of the emitter with decay time $T_1$ as well as pure dephasing processes with dephasing time $T_2$. This is valid when the 2-photon decay dominates, i.e., $\Gamma\gg T_1^{-1},T_2^{-1}$. In this section we show that for the parameters estimated at the end of the main text, this is indeed a very good approximation. To do so we consider the case of $N=2$ emitters and simulate the master equation
\begin{equation}
\dot{\rho}=\frac{\Gamma}{2}\left(2S_{-}^{2}\rho S_{+}^{2}-S_{+}^{2}S_{-}^{2}\rho-\rho S_{+}^{2}S_{-}^{2}\right)+\frac{1}{2T_1}\sum_{i=1}^{2}\left(2\sigma_{+}^{(i)}\rho\sigma_{-}^{(i)}-\sigma_{+}^{(i)}\sigma_{-}^{(i)}\rho-\rho\sigma_{+}^{(i)}\sigma_{-}^{(i)}\right)
+\frac{1}{2T_2}\sum_{i=1}^{2}\left(\sigma_{z}^{(i)}\rho\sigma_{z}^{(i)}-\rho\right),
\end{equation}
where $\Gamma$ is the rate for the two-photon decay process as defined in the main text. In Fig.~\ref{masterequation} we plot the solutions of this equation for $J/(2\pi)=100$ MHz, $\Gamma/J\approx 5 \times 10^{-4}$ and $T_{1}=15\,\mu$s, $ T_2 = 7.5\,\mu $s. The value of $\Gamma$ is obtained for $U/J=1$ and by choosing the other waveguide parameters as in Fig.~\ref{dynamics}(a) green solid line. For these parameters we find almost no difference in the decay rate, but during the decay the singly-excited states become populated, which leads to a slightly slower decay at the end of the process.}

\section{Semi-analytical results for subradiance}

In the main text, we discussed subradiance with $N=4$ emitters.
In this section, we derive the semi-analytical expressions for the
transition rates shown in Fig.~3(b) and the long time limit of the excited
state population $P_e(t)$ in Fig.~3(a) of the main text.


The two-excitation subspace for the $N=4$ problem is spanned by six states
\begin{align}
|eegg\rangle,&&|egeg\rangle,&&|egge\rangle,&&|geeg\rangle,&&|gege\rangle,&&|ggee\rangle.
\end{align}
Defining the 6 eigenstates of $H_{\rm eff}$ in this subspace
as $|T_i\rangle\,$, the transition rate from the state $|T_i\rangle$ to
the ground state $|G\rangle=|gggg\rangle$ is $R_i^{G}=\sum_m w_m|\langle T_i|Y_m\rangle|^2$, with $w_m$ and $|Y_m\rangle$ being the eigenvalue and the corresponding eigenvector
of the decay matrix $\text{Re}\, A_{i',j'}=\text{Re}\,A_{ij,kl}$.
Following a similar approach, we can obtain the transition rate from the excited state
$|E\rangle$ to the two-excitation state $|T_i\rangle$ as $R_i^{E}=\langle T_i|\rho_E|T_i\rangle$, where $\rho_E=\sum_m w_m |\bar{Y}_m\rangle\langle \bar{Y}_m|/\mathcal{N}$. We have defined $\mathcal{N}=\sum_m{w_m}$ and the state $|\bar{Y}_m\rangle=\bigotimes_{i=1}^4\sigma_i^x|Y_m\rangle$, i.e.~$|\bar Y_m\rangle$ is the state $|Y_m\rangle$ but with the swap $g\leftrightarrow e$ made for all the emitters. We should note that
$|\bar{Y}_m\rangle\neq|Y_m\rangle$, so it is possible to have different rates for decaying
into and out of the states in two-excitation subspace.

In the case examined in the main text, with the emitters evenly spaced, we find one exact dark state  $|D_2\rangle$ with $R_{D_2}^E\neq0$ but $R_{D_2}^G=0$, i.e.\ the dynamics decays into this state but then cannot escape. The excited state
 population after a long time evolution can be expressed as $P_e(t=\infty)=R_{D_2}^{E}/2$.

To further explain the nature of the dark state, we first note that this state only has support in the subspace spanned by the states $\{|egge\rangle,|geeg\rangle\}$. In this 2D space the non-Hermitian Hamiltonian $H_{\rm eff}$
can be expressed as
\begin{equation}
H_{\rm eff}=-4i\Gamma_{0}\left(\begin{array}{cc}
f_{K_{0}}^{2}(3x) & f_{K_{0}}(x)f_{K_{0}}(3x)\\
f_{K_{0}}(x)f_{K_{0}}(3x) & f_{K_{0}}^{2}(x)
\end{array}\right).
\end{equation}
 So that, $|D_2\rangle$ is then the eigenstate corresponding to the zero eigenvalue of this matrix and is given by
 \begin{equation}
 |D_{2}\rangle=\alpha(x)|egge\rangle-\beta(x)|geeg\rangle,
 \end{equation}
 where
 \begin{equation}
 \frac{\alpha(x)}{\beta(x)}=\frac{f_{K_{0}}(x)}{f_{K_{0}}(3x)} \geq 1.
 \end{equation}

We also find that the matrix $\text{Re}\,A$ is only rank-$2$, and hence in the above analysis only 2 of the $w_m$ are non-zero. Therefore, the state reached from the fully excited state, $|E\rangle$,  can be expressed as
\begin{equation}
\rho_E=\frac{w_1|\bar{Y}_1\rangle\langle \bar{Y}_1|+w_2|\bar{Y}_2\rangle\langle \bar{Y}_2|}{w_1+w_2}.
\end{equation}
It can be shown that one of these states is orthogonal to the dark state $\langle D_2|
\bar{Y}_{2}\rangle=0$. Therefore,
the final result for $P_e(t=\infty)$ can be expressed in the simple form
\begin{equation}
P_e(t=\infty)=\frac{w_1|\langle D_2|\bar{Y}_1\rangle|^2}
{2(w_1+w_2)},
\end{equation}
which is the result shown by the dashed lines in Fig.~3(a) of the main text.

\section{Super-correlated emission}
In the main text, we discuss the two-photon decay process when all of the TLSs are located in the same site of the waveguide, which gives rise to super-correlated emission. Here, we will give a comparison between the mean-field (MF) approximation and the numerical integration of the master equation for both the super-correlated emission and conventional superradiance.

\begin{figure}
\begin{centering}
\includegraphics[width=0.95\columnwidth]{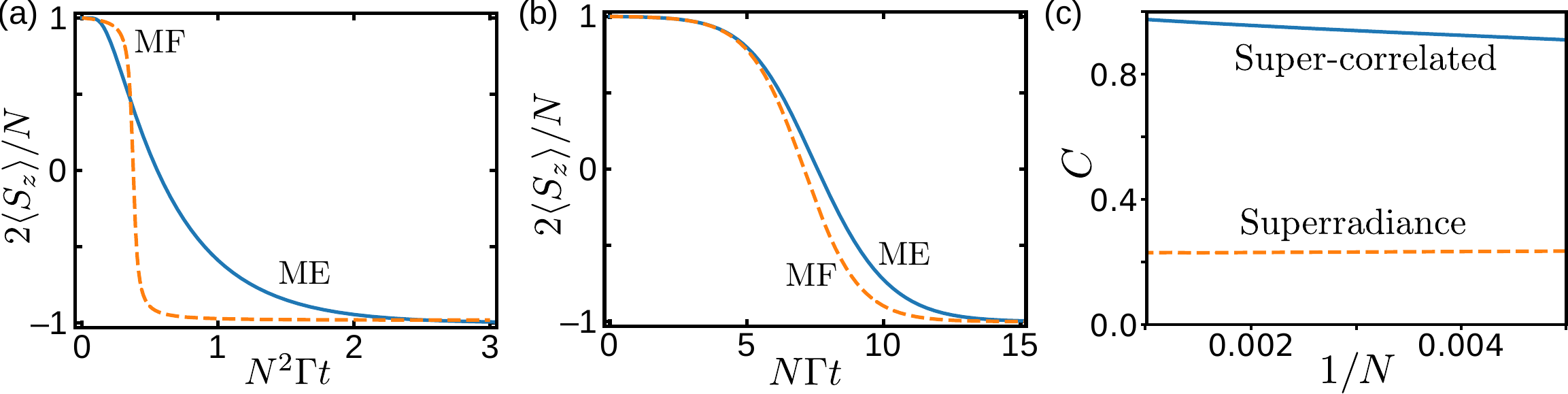}
\end{centering}
\caption{The dynamics of $\langle S_z\rangle$ for super-correlated emission (a) and conventional superradiance (b). The solid lines are the results of solving the master equation, the dashed lines are the mean-field results. (c) The correlation parameter as defined in the text for the two cases. In (a) and (b) we show results for $N=1200$. }
\label{fig:SR}
\end{figure}

For the super-correlated emission, the master equation is given by
\begin{equation}
\dot \rho=\frac{\Gamma}{2}\left(2S_{-}^{2}\rho S_{+}^{2}-\rho S_{+}^{2}S_{-}^{2}-S_{+}^{2}S_{-}^{2}\rho\right).
\label{masters}
\end{equation}
Under the MF approximation, where $\langle S_z^2\rangle\approx\langle S_z\rangle^2$, the dynamics of $\langle S_z\rangle$ can be described by
\begin{equation}
\begin{split}
\frac{d}{dt}\langle S_{z}\rangle&=-2\Gamma\langle S_{+}^{2}S_{-}^{2}\rangle
= \langle S_+S_- \left[ S_+S_- + 2\left( S_z - 1 \right) \right] \rangle\\
&\approx-2\Gamma[S(S+1) - \langle S_z\rangle^2 + \langle S_z \rangle][S(S+1) - \langle S_z\rangle^2 + 3\langle S_z \rangle - 2].
\end{split}
\end{equation}
where we have used $ \left[ S_z, S_- \right] = -S_- $, $ \left[ S_+, S_- \right] = 2S_z $ and $ S_+S_- = \vec S^2 - S_z^2 + S_z $. In contrast, for conventional superradiance, the master equation is
\begin{equation}
\dot \rho=\frac{\Gamma}{2}\left(2S_{-}\rho S_{+}-\rho S_{+}S_{-}-S_{+}S_{-}\rho\right),
\label{masters1}
\end{equation}
and as a result, we obtain for the MF equation \cite{Gross1982}
\begin{equation}
\frac{d}{dt}\langle S_{z}\rangle=-\Gamma\langle S_{+}S_{-}\rangle
\approx-\Gamma[S(S+1)-\langle S_z\rangle^2+\langle S_z\rangle].
\end{equation}
In Fig.~\ref{fig:SR}(a), we give the comparison between the MF approximation and exact numerical results based on the master equation for the super-correlated emission for the initial state as $|\psi(0)\rangle=|S_z=S\rangle$. Equivalent results for conventional superradiance are shown in Fig.~\ref{fig:SR}(b). It clearly demonstrates that the MF approximation works well for conventional superradiance, but breaks down for super-correlated emission, even in the limit of a large number of atoms, $N=1200$, as shown here.

{The qualitative difference between these two sets of results can be quantified by the correlation parameter
\begin{equation}
C = {\max_{t \in [0,\infty)}} \frac{4\left(\langle S_z^2\rangle(t)-\langle S_z(t)\rangle^2\right)}{N^2},
\end{equation}
as defined in the main text.  We show $C$  in Fig.~\ref{fig:SR}(c) for different numbers of emitters $N$. We see that in the limit $N\to\infty$ $C\approx 0.2$ for superradiance and $C\approx 1-O(1/N)$ for the super-correlated decay process, showing the different way in which the decay process happens in these two models.}

\section{Single-cavity implementation}

\begin{figure}
\begin{centering}
\includegraphics[width=0.9\columnwidth]{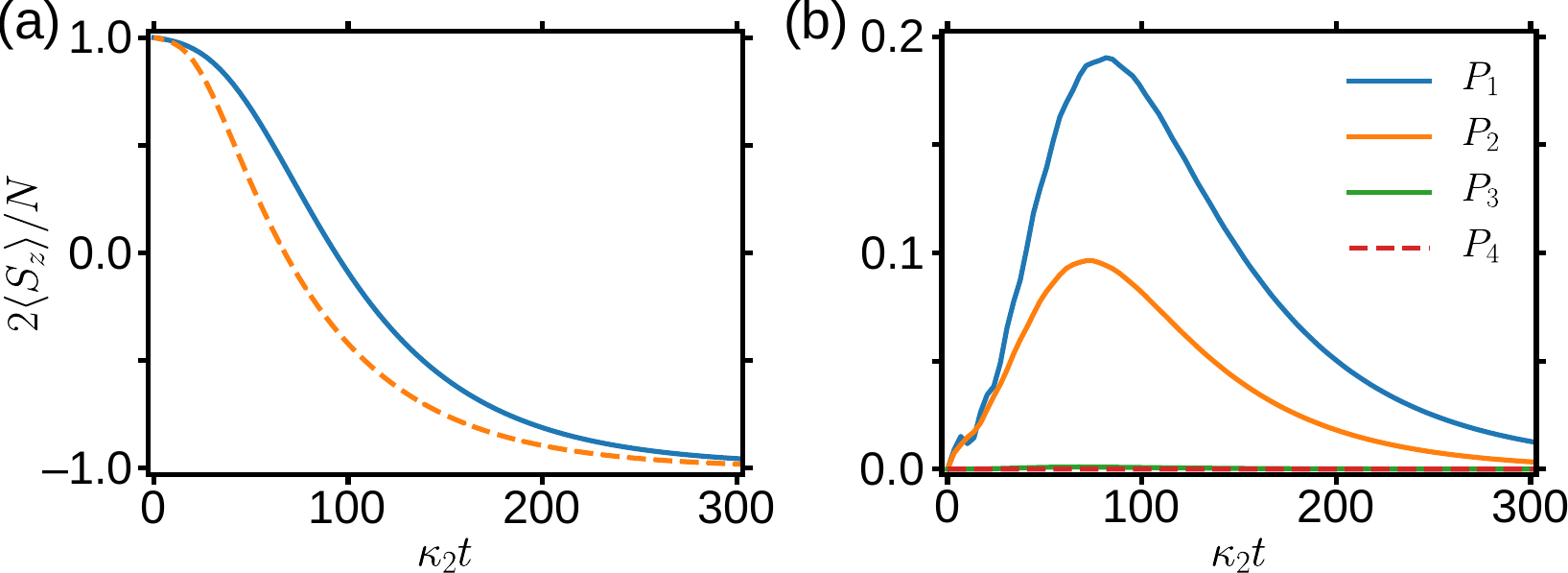}
\end{centering}
\caption{{The dynamics of an ensemble of emitters coupled to a single nonlinear cavity with nonlinear decay (a) and the photon number populations (b). In (a) the solid blue line shows the results obtained from the full model in Eq.~\eqref{eq:SingleCavityExact}, which includes the dynamics of the photons. The orange dashed line shows the results obtained from the effective master equation, Eq.~\eqref{effmaster}. In both plots the parameters are set as $\kappa_{1}=0.01\kappa_{2}$, $U=30\kappa_{2}$, $g=0.12\kappa_{2}$ and $\Delta=U/2$. We have considered $N = 50$ emitters and in the exact model we have included up to $n_{\rm ph}=4$ photons.}}
\label{cavitymodel}
\end{figure}

{
In the above sections we have discussed the super-correlated emission of atoms located in a waveguide. In this section we show how similar physics can be observed in a single-cavity realization. To show this, we consider a model consisting of an ensemble of two-level atoms interacting with
a nonlinear cavity. The cavity has engineered two-photon decay~\cite{leghtas2015}, at a rate $\kappa_2$, that exceeds the single photon decay rate, $\kappa_1$. The Hamiltonian of the whole system is
\begin{equation}
H_{SC}=\omega_{c}a^{\dagger}a-\frac{U}{2}a^{\dagger}a^{\dagger}aa+\omega_{e}S_{z}+g(a^{\dagger}S_{-}+aS_{+}),
\label{Har}
\end{equation}
where $a$ is the annihilation operator of the cavity mode and $g$ is cavity-emitter coupling strength. Similar to the waveguide setup, we adiabatically eliminate the single-photon states and obtain the effective Hamiltonian
\begin{equation}
H_{\rm eff}=(2\omega_{c}-U)B^{\dagger}B+\omega_{e}S_{z}+G(BS_{+}^{2}+B^{\dagger}S_{-}^{2}),
\end{equation}
which describes the emitter-field coupling via a two-photon process. Here, assuming that the cavity remains close to the vacuum state, $B=|0\rangle\langle2|$, the effective coupling strength is $G=\sqrt{2}g^{2}/\Delta$ and $ \Delta = \omega_c - \omega_e $. When the two-photon decay rate, $\kappa_2$, of the cavity is much larger than the effective coupling, $G$, the cavity acts as a Markovian reservoir for the ensemble of emitters and we can again derive a Born-Markov master equation for the reduced density matrix of the emitters only. We obtain
\begin{equation}
\dot{\rho}=A(S_{-}^{2}\rho S_{+}^{2}-S_{+}^{2}S_{-}^{2}\rho)+A^{*}(S_{-}^{2}\rho S_{+}^{2}-\rho S_{+}^{2}S_{-}^{2}),
\label{effmaster}
\end{equation}
where
\begin{eqnarray}
A=G^{2}\int_{0}^{\infty}d\tau\langle B(\tau)B^{\dagger}(0)\rangle
e^{2i\omega_{a}\tau}\nonumber =G^{2}\int_{0}^{\infty}d\tau e^{[2i(\omega_{a}-\omega_{c}+\frac{U}{2})-\kappa_{2}]\tau}\nonumber =-\frac{G^{2}}{2i(\omega_{e}-\omega_{c}+\frac{U}{2})-\kappa_{2}}.
\end{eqnarray}
When the parameters satisfy $\omega_e=\omega_c-U/2$, that is, two emitters are resonant with the two-photon transition, the collective two-photon emission rate simplifies to
\begin{equation}
A=\frac{G^{2}}{\kappa_{2}}=\frac{8g^{4}}{U^{2}\kappa_{2}}.
\end{equation}
We see that in this case the master equation reduces to the master equation for the collective two-photon decay, Eq.~(12) in the main text.

To check the validity of the effective master equation in this single-cavity configuration, we plot in Fig.~\ref{cavitymodel}(a) the dynamics of $\langle S_z\rangle$ based on Eq.~\eqref{effmaster}. This result is compared to the exact dynamics of $\langle S_z\rangle$ when the cavity mode is fully taken into account. The exact dynamics includes two-photon and single-photon decay and is described by the master equation
\begin{equation}\label{eq:SingleCavityExact}
\dot{\rho}=-i[H_{SC},\rho]+\frac{\kappa_{2}}{2}\left(2a^{2}\rho a^{\dagger2}-a^{\dagger2}a^{2}\rho-\rho a^{\dagger2}a^{2}\right) +\frac{\kappa_{1}}{2}\left(2a\rho a^{\dagger}-a^{\dagger}a\rho-\rho a^{\dagger}a\right),
\end{equation}
where $\kappa_1(\ll\kappa_2)$ is the single-photon decay rate of the cavity mode and the Hamiltonian $H_{SC}$ is given in Eq.~\eqref{Har}.

Due to the symmetry of the problem, the model can be restricted to collective spin states with maximal spin quantum number. Compared to the waveguide this allows us to simulate much larger numbers of emitters and include states with up to $n_{\rm ph}=4$ photons in the cavity. The results in Fig.~\ref{cavitymodel}(a) and Fig.~\ref{cavitymodel}(b) show that for a realistic set of parameters and up to $N=50$ the exact dynamics is very accurately reproduced by the effective master equation and that states with more than $n_{\rm ph}=2$ photons in the cavity are hardly populated. The dominante correction comes from the state with $n_{\rm ph}=1$ due to a finite hybridization between emitters and photons. However, this state has a slow decay and doesn't affect considerably the dynamics.

For a much larger number of emitters the Markov approximation eventually breaks down and we observe stronger differences between the effective and the exact model. Empirically, we find that the condition
\begin{equation}
 \Gamma N^3 \lesssim \kappa_2
\end{equation}
should be satisfied for the validity of the effective master equation. For the original waveguide scenario we expect a similar condition to hold when the 2-photon decay rate $\kappa_2$ is replaced by the group velocity $v_g(K_0)$.}

\begin{thebibliography}{99}
\bibitem{Dicke} R. H. Dicke, {\it Coherence in spontaneous radiation processes}, Phys. Rev. {\bf 93}, 99 (1954).

\bibitem{Gross1982main} M. Gross and S. Haroche, {\it Superradiance: An essay on the theory of collective
spontaneous emission}, Phys. Rep. {\bf 93}, 301 (1982). 

\bibitem{Hughes2004} S. Hughes, {\it Enhanced single-photon emission from quantum dots in photonic crystal waveguides and nanocavities}, Opt. Lett. {\bf 29}, 2659 (2004).

\bibitem{Shen2005} J.-T. Shen and S. Fan, {\it  Coherent single photon transport in a one-dimensional waveguide coupled with superconducting quantum bits},  Phys. Rev. Lett. {\bf 95}, 213001 (2005).

\bibitem{Chang2007} D. E. Chang, A. S. S{\o}rensen, E. A. Demler, and M. D. Lukin, {\it A single-photon transistor using nanoscale surface plasmons}, Nature Phys. {\bf 3}, 807 (2007).

\bibitem{Zhou2008} L. Zhou,  Z. R. Gong, Y.-x. Liu, C. P. Sun, and F. Nori, {\it Controllable scattering of a single photon inside a one-dimensional resonator waveguide}, Phys. Rev. Lett. {\bf 101}, 100501 (2008).

\bibitem{Longo2010} P. Longo, P. Schmitteckert, and K. Busch, {\it Few-photon transport in low-dimensional systems: interaction-induced radiation trapping}, Phys. Rev. Lett. {\bf 104}, 023602 (2010).

\bibitem{Zheng2010} H. Zheng, D. J. Gauthier, and H. U. Baranger, {\it Waveguide QED: Many-body bound-state effects in coherent and Fock-state scattering from a two-level system}, Phys. Rev. A {\bf 82}, 063816 (2010).



\bibitem{Lombardo2014} F. Lombardo, F. Ciccarello, and G. M. Palma, {\it Photon localization versus population trapping in a coupled-cavity array}, Phys. Rev. A {\bf 89}, 053826 (2014).  

 \bibitem{Shahmoon2016} E. Shahmoon, P. Grisins, H. P. Stimming, I. Mazets, and G. Kurizki, {\it Highly nonlocal optical nonlinearities in atoms trapped near a waveguide}, Optica {\bf 3}, 725 (2016).


\bibitem{Roy2017} D. Roy. C. M. Wilson, and O. Firstenberg, {\it Colloquium: Strongly interacting photons in one-dimensional continuum}, Rev. Mod. Phys. {\bf 89}, 021001 (2017).



\bibitem{ReitzPRL2013} D. Reitz, C. Sayrin, R. Mitsch, P. Schneeweiss, and A. Rauschenbeutel, \textit{Coherence properties of nanofiber-trapped Cesium atoms}, Phys. Rev. Lett. {\bf 110}, 243603 (2013). 

\bibitem{YallaPRL2014} R. Yalla, M. Sadgrove, K. P. Nayak, and K. Hakuta,\textit{ Cavity quantum electrodynamics on a nanofiber using a composite photonic crystal cavity}, Phys. Rev. Lett. {\bf 113}, 143601 (2014). 

\bibitem{Hood2016} J. D. Hood, {\it et al}., {\it Atom-atom interactions around the band edge of a photonic crystal waveguide}, PNAS {\bf 113}, 10507 (2016). 

\bibitem{Corzo2019} N. V. Corzo, J. Raskop, A. Chandra, A. S. Sheremet, B. Gouraud, and J. Laurat, {\it Waveguide-coupled single collective excitation of atomic arrays},  Nature {\bf 566}, 359 (2019).



\bibitem{review-lodahl} P. Lodahl, S. Mahmoodian, and S. Stobbe, \textit{Interfacing single photons and single quantum dots with photonic nanostructures}, Rev. Mod. Phys. {\bf 87}, 347 (2015).


\bibitem{Astafiev2010} O. Astafiev, {\it et al}., {\it Resonance fluorescence
of a single artificial atom}, Science {\bf 327}, 840 (2010).

\bibitem{Hoi2011} I.-C. Hoi, C. M. Wilson, G. Johansson, T. Palomaki, B. Peropadre, and P.
Delsing, {\it Demonstration of a single-photon router in the microwave regime},
Phys. Rev. Lett. {\bf 107}, 073601 (2011).

\bibitem{Mlynek2014} J. A. Mlynek, A. A. Abdumalikov, C. Eichler, and A. Wallraff, {\it Observation of Dicke superradiance for two artificial atoms in a cavity with high decay rate}, Nat. Commun. {\bf 5}, 5186 (2014).


\bibitem{Mirhosseini2018} M. Mirhosseini, {\it at al}., {\it Cavity quantum electrodynamics with atom-like mirrors}, Nature {\bf 569}, 692 (2019).

\bibitem{Sundaresan2019} N. M. Sundaresan, R. Lundgren, G. Zhu, A. V. Gorshkov, and A. A. Houck, {\it Interacting qubit-photon bound states with superconducting circuits}, Phys. Rev. X {\bf 9}, 011021 (2019).



\bibitem{ChangPRL2013} D. E. Chang, J. I. Cirac, and H. J. Kimble, {\it Self-organization of atoms along a nanophotonic waveguide}, Phys. Rev. Lett. {\bf 110}, 113606 (2013).
%
\bibitem{GriesserPRL2013}  T. Grie\ss er and H. Ritsch, {\it Light-induced crystallization of cold atoms in a 1D optical trap}, Phys. Rev. Lett. {\bf 111}, 055702 (2013).

\bibitem{dzsotjan2010} D. Dzsotjan, A. S. S{\o}rensen, and M. Fleischauer, \textit{Quantum emitters coupled to surface plasmons of a nanowire: A Green's function approach}, Phys. Rev. B \textbf{82}, 075427 (2010).

\bibitem{GonzalesTudela2011} A. Gonzalez-Tudela, D. Martin-Cano, E. Moreno, L. Martin-Moreno, C. Tejedor, and F. J. Garcia-Vidal, {\it Entanglement of two qubits mediated by one-dimensional plasmonic waveguides}, Phys. Rev. Lett. {\bf 106}, 020501 (2011).

\bibitem{Stannigel2012} K. Stannigel, P. Rabl, and P. Zoller, {\it Driven-dissipative preparation of entangled
states in cascaded quantum-optical networks}, New J. Phys. {\bf 14}, 063014 (2012).

\bibitem{Zheng2013} H. Zheng and H. U. Baranger, {\it Persistent quantum beats and long-distance entanglement from waveguide-mediated interactions}, Phys. Rev. Lett. {\bf 110}, 113601 (2013).

\bibitem{shahmoon2013} E. Shahmoon and G. Kurizki, \textit{Nonradiative interaction and entanglement between distant atoms}, Phys. Rev. A \textbf{87}, 033831 (2013).

\bibitem{Facchi2016} P. Facchi, M. S. Kim, S. Pascazio, F. V. Pepe, D. Pomarico, and T. Tufarelli, {\it Bound states and entanglement generation in waveguide quantum electrodynamics}, Phys. Rev. A {\bf 94}, 043839 (2016).



\bibitem{Mahmoodian2016} S. Mahmoodian, P. Lodahl, and A. S. S{\o}rensen, {\it Quantum networks with chiral light-matter interaction in waveguides}, Phys. Rev. Lett. {\bf 117}, 240501 (2016).

\bibitem{AsenjoGarcia2017} A. Asenjo-Garcia, M. Moreno-Cardoner, A. Albrecht, H. J. Kimble, and D. E. Chang, {\it Exponential improvement in photon storage fidelities using subradiance and ``selective radiance" in atomic arrays}, Phys. Rev. X {\bf 7}, 031024 (2017).






\bibitem{Hartmann2006} M. J. Hartmann, F. G. S. L. Brand\~ao and M. B. Plenio, \textit{Strongly interacting polaritons in coupled arrays of cavities}, Nature Phys. {\bfseries 2}, 849 (2006).
%
\bibitem{GreentreeNatPhys2006} A. D. Greentree, C. Tahan, J. H. Cole, and C. L. Hollenberg, \textit{Quantum phase transitions of light}, Nature Phys. {\bf 2}, 856 (2006).
%
\bibitem{AngelakisPRA2007} D. G. Angelakis, M. F. Santos, and S. Bose, \textit{Photon-blockade-induced Mott transitions and XY spin models in coupled cavity arrays},  Phys. Rev. A {\bf 76}, 031805(R) (2007).

 \bibitem{Carusotto2013} I. Carusotto and C. Ciuti, {\it Quantum fluids of light}, Rev. Mod. Phys. {\bf 85}, 299 (2013).

\bibitem{Angelakis2017} D. G. Angelakis, \textit{Quantum Simulations with Photons and Polaritons},
(Springer, New York, 2017).

\bibitem{Houck2012} A. A. Houck, H. E. T\"{u}reci, and J. Koch, {\it On-chip quantum simulation with superconducting circuits}, Nature Physics {\bf 8}, 292 (2012).

\bibitem{Degiron2010} A. Degiron and D. R. Smith, {\it Nonlinear long-range plasmonic waveguides}, Phys. Rev. A {\bf 82}, 033812 (2010).

\bibitem{Kauranen2012} M. Kauranen and A. V. Zayats, {\it Nonlinear plasmonics}, Nature Photonics {\bf 6}, 737 (2012).


\bibitem{SHbook} S. Haroche and J.-M. Raimond, {\it Exploring the Quantum: Atoms, Cavities, and Photons}, (Oxford University Press, New York, 2006).

\bibitem{Black2005} A. T. Black, J. K. Thompson, and V. Vuleti\'{c}, {\it On-demand superradiant conversion of atomic spin gratings into single photons with high efficiency},  Phys. Rev. Lett. {\bf 95}, 133601 (2005).

\bibitem{Akkermans2008} E. Akkermans, A. Gero, and R. Kaiser, {\it Photon localization and Dicke superradiance in atomic gases}, Phys. Rev. Lett. {\bf 101}, 103602 (2008).

\bibitem{Scully2009} M. O. Scully, {\it Collective Lamb shift in single photon Dicke superradiance},  Phys. Rev. Lett. {\bf 102}, 143601 (2009).

\bibitem{Bienaime2012} T. Bienaim\'{e}, N. Piovella, and R. Kaiser, {\it Controlled Dicke subradiance from a large cloud of two-level systems}, Phys. Rev. Lett. {\bf 108}, 123602 (2012).

\bibitem{Ostermann2013} L. Ostermann, H. Ritsch, and C. Genes, {\it Protected state enhanced quantum metrology with interacting two-level ensembles}, Phys. Rev. Lett. {\bf 111}, 123601 (2013).

\bibitem{Scully2015} M. O. Scully, {\it Single photon subradiance: quantum control of spontaneous emission and ultrafast readout}, Phys. Rev. Lett. {\bf 115}, 243602 (2015). 

\bibitem{Angerer2018} A. Angerer, {\it et al}., {\it Superradiant hybrid quantum devices}, Nature Physics {\bf 14}, 1168 (2018).

\bibitem{Zhang2019} Y.-X. Zhang and K. M\o lmer, {\it Theory of subradiant states of a one-dimensional two-level atom chain}, Phys. Rev. Lett. {\bf 122}, 203605 (2019).

\bibitem{Ke2019} Y. Ke, A. V. Poshakinskiy, C. Lee, Y. S. Kivshar, and A. N. Poddubny, {\it Inelastic scattering of photon pairs in qubit arrays with subradiant states}, Phys. Rev. Lett. {\bf 123}, 253601 (2019).











\bibitem{supp} See the Supplementary Material, which includes the additional reference~\cite{leghtas2015}, for a discussion of the nature of the bound states, derivation of the master equation used, more details about the subradiant states and mean-field results about the super-correlated decay process.


\bibitem{leghtas2015} Z. Leghtas, {\it et al}., \textit{Confining the state of light to a quantum manifold by engineered two-photon loss}, Science \textbf{347}, 853 (2015).


\bibitem{Piil2007} R. Piil and K. M\o lmer, {\it Tunneling couplings in discrete lattices, single-particle band structure, and eigenstates of interacting atom pairs}, Phys. Rev. A {\bf 76}, 023607 (2007).

\bibitem{Valiente2008} M. Valiente and D. Petrosyan, {\it Two-particle states in the Hubbard model}, J. Phys. B: At. Mol. Opt. Phys. {\bf 41}, 161002 (2008).

\bibitem{Winkler2004} K. Winkler, {\it et al}., {\it Repulsively bound atom pairs in an optical lattice}, Nature  {\bf 441}, 853 (2006).

\bibitem{Calajo2016} G. Calaj\'{o}, F. Ciccarello, D. Chang, and P. Rabl, {\it Atom-field dressed states in slow-light waveguide QED}, Phys. Rev. A {\bf 93}, 033833 (2016).

\bibitem{BreuerBook} H. Breuer and F. Petruccione, \textit{The Theory of Open Quantum
Systems} (Oxford University Press, Oxford, UK, 2002).


\bibitem{Chang2012} D. E. Chang, L. Jiang, A. V. Gorshkov, and H. J. Kimble, {\it Cavity QED with atomic mirrors}, New J. Phys. {\bf 14}, 063003 (2012). 

\bibitem{GonzalesTudela2013} A. Gonzales-Tudela and D. Porras, {\it Mesoscopic entanglement induced by spontaneous emission in solid-state quantum optics}, Phys. Rev. Lett. {\bf 110}, 080502 (2013). 


\bibitem{Daley2014} A. J. Daley,  {\it Quantum trajectories and open many-body quantum
systems}, Advances in Physics {\bf 63}, 77 (2014).


\bibitem{Leib2012} M. Leib, F. Deppe, A. Marx, R. Gross, and M. J. Hartmann, {\it Networks of nonlinear superconducting transmission line resonators}, New J. Phys. {\bf 14}, 075024 (2012). 

\bibitem{HacohenGourgy2015} S. Hacohen-Gourgy, V. V. Ramasesh, C. D. Grandi, I. Siddiqi, and S. M. Girvin, {\it Cooling and autonomous feedback in a Bose-Hubbard chain with attractive interactions}, Phys. Rev. Lett. {\bf 115}, 240501 (2015).

\bibitem{Roushan2017} P. Roushan, \textit{et al.}, {\it Spectroscopic signatures of
localization with interacting photons in superconducting qubits}, Science {\bf 358}, 1175 (2017).

\bibitem{Ma2019} R. Ma, {\it et al}., {\it A dissipatively stabilized Mott insulator of photons}, Nature {\bf 566}, 51 (2019). 


\bibitem{Kjaergaard2019} M. Kjaergaard, {\it et al}., {\it Superconducting qubits: current state of play}, Annual Review of Condensed Matter Physics {\bf 11}, 369 (2020).


\bibitem{Petrosyan2008} D. Petrosyan and M. Fleischhauer,  \textit{Quantum information processing with single photons and atomic ensembles in microwave coplanar waveguide Resonators}, Phys. Rev. Lett. {\bf100}, 170501 (2008).

\bibitem{Hogan2012} S. D. Hogan, J. A. Agner, F. Merkt, T. Thiele, S. Filipp, and A. Wallraff,  \textit{Driving Rydberg-Rydberg transitions from a coplanar microwave Waveguide}, Phys. Rev. Lett. {\bf 108}, 063004 (2012).

\bibitem{Beck2016} M. A. Beck, {\it et al}., {\it Optimized coplanar waveguide resonators for a superconductor-atom interface},  Appl. Phys. Lett. {\bf 109}, 092602 (2016).

\bibitem{Calajo2017} G. Calaj\'{o} and P. Rabl, {\it Strong coupling between atoms and slow-light Cherenkov photons}, Phys. Rev. A {\bf 95}, 043824 (2017).






\end{thebibliography}

\begin{thebibliography}{99}
\bibitem{Piil2007} R. Piil and K. M\o lmer, {\it Tunneling couplings in discrete lattices, single-particle band structure, and eigenstates of interacting atom pairs}, Phys. Rev. A {\bf 76}, 023607 (2007).


\bibitem{Winkler2004} K. Winkler, G. Thalhammer, F. Lang, R. Grimm, J. Hecker Denschlag, A. J. Daley, A. Kantian, H. P. B\"uchler, and P. Zoller, {\it Repulsively bound atom pairs in an optical lattice}, Nature  {\bf 441}, 853 (2006).
		
\bibitem{GC} G. Calaj\'{o}, F. Ciccarello, D. Chang, and P. Rabl, {\it Atom-field dressed states in slow-light waveguide QED}, Phys. Rev. A {\bf 93}, 033833 (2016).


\bibitem{HB}H. Breuer and F. Petruccione, \textit{The Theory of Open Quantum
Systems} (Oxford University Press, Oxford, UK, 2002).

\bibitem{Gross1982} M. Gross and S. Haroche, {\it Superradiance: An essay on the theory of collective spontaneous emission}, Phys. Rep. {\bf 93}, 301 (1982). 

\bibitem{leghtas2015} Z. Leghtas, S. Touzard, I. M. Pop1, A. Kou, B. Vlastakis, A. Petrenko, K. M. Sliwa, A. Narla, S. Shankar, M. J. Hatridge, M. Reagor, L. Frunzio, R. J. Schoelkopf, M. Mirrahimi, and M. H. Devoret, \textit{Confining the state of light to a quantum manifold by engineered two-photon loss}, Science \textbf{347}, 853 (2015).

\end{thebibliography}
\end{document}